\documentclass[twocolumn]{aastex61}

\usepackage{amsmath,amssymb}
\usepackage{graphicx}
\usepackage{cases}
\usepackage{bm}

\newcommand{\Y}{Y_{lm}(\theta , \phi)}
\newcommand{\e}{\mathrm{e}}
\newcommand{\diff}{\mathrm{d}}

\newcommand{\ltsim}{\protect\raisebox{-0.5ex}{$\:\stackrel{\textstyle <}{\sim}\:$}} 
\newcommand{\gtsim}{\protect\raisebox{-0.5ex}{$\:\stackrel{\textstyle >}{\sim}\:$}} 

\newcommand{\E}[1]{\times 10^{#1}}

\accepted{February 19, 2019}
\submitjournal{ApJ}

\shorttitle{Linear analysis of the shock instability in core-collapse supernovae}
\shortauthors{Sugiura et al.}

\begin{document}

\title{LINEAR ANALYSIS OF THE SHOCK INSTABILITY IN CORE-COLLAPSE SUPERNOVAE: \\INFLUENCES OF ACOUSTIC POWER AND FLUCTUATIONS OF NEUTRINO LUMINOSITY}

\correspondingauthor{Ken'ichi Sugiura}
\email{sugiura@heap.phys.waseda.ac.jp}

\author{KEN'ICHI SUGIURA}
\affiliation{Advanced Research Institute for Science and Engineering, Waseda University, 3-4-1 Ookubo, Shinjuku-ku, Tokyo, 169-8555, Japan}

\author{KAZUYA TAKAHASHI}
\affiliation{Center for Gravitational Physics, Yukawa Institute for Theoretical Physics, Kyoto University, Kyoto 606-8502, Japan}

\author{SHOICHI YAMADA}
\affiliation{Science and Engineering, Waseda University, 3-4-1 Ookubo, Shinjuku-ku, Tokyo, 169-8555, Japan}

\begin{abstract}
This paper is a sequel to \citet{Taka16}, in which the authors investigated the influences of fluctuations in pre-shock accreting matter on the linear stability of the standing accretion shock in core-collapse supernovae (CCSNe).
If one understands that this was concerning the effect of the outer boundary condition for the post-shock accretion flows, the present paper should be regarded as an investigation on the possible influences of the inner boundary conditions. More specifically, we impose a time-dependent, oscillating condition at the inner boundary, having in mind the injection of acoustic power by an oscillating proto-neutron star. We also consider possible correlations between the inner and outer boundary conditions as invoked in the argument for Lepton-number Emission Self-sustained Asymmetry, or LESA.
In this paper, we conduct the linear stability analysis of the standing accretion shock commonly encountered in CCSNe based on Laplace transform. We find that the acoustic power enhances the standing accretion shock instability, or SASI, especially when the luminosity is low.
On the other hand, the correlation between the fluctuations of neutrino luminosity at the neutrino sphere has little influences on the instability, changing the amplitudes of eigenmodes only slightly.
We further investigate steady solution of perturbation equations, being motivated by LESA, and conclude that not the difference but the sum of the fluxes of $\nu_e$ and $\bar{\nu}_e$ is the key ingredient to production of the self-sustained steady perturbed configuration.
\end{abstract}

\keywords{instabilities---methods: analytical --- supernovae: general}

\section{Introduction} \label{sec:intro}
Core-collapse supernovae (CCSNe) are explosions in the final stage of massive star evolutions. 
Comprehension of the CCSNe mechanism is important not only for its own sake but also for understanding the formation of neutron stars or black holes as well as the synthesis of heavy elements. 
Although a remarkable progress has been made over the years, the mechanism of CCSNe has not yet been fully understood (e.g. see \citet{Janka17b}). 
It is well known, however, that the shock wave produced at core bounce does not propagate through the entire core promptly but is stagnated somewhere inside it because of energy losses via  photodissociations and hence there should be some mechanism to push the stalled shock wave outward again. 

The most promising scenario at present is the neutrino-heating mechanism, in which matter passing the stalled shock wave acquires energy from neutrinos emitted from a proto-neutron star (PNS) and the shock revival obtains as a result. It is now a consensus of the supernova society that multidimensional effects are crucial for the success of this scenario except for the low-mass end of massive stars (\citet{Kitaura06}). In fact, it is believed that CCSNe are not spherically symmetric in general, being subject to hydrodynamical instabilities (see \citet{Burrows13,Janka16} for recent reviews) such as convection and standing accretion shock instability (SASI) (\citet{Bethe90, Herant94, Blondin03, Iwakami08}). These instabilities induce turbulent motions and extra pressure provided by the turbulence pushes the shock wave outward and, as a result, the gain region is broadened and the neutrino heating is enhanced (\citet{Murphy13, Couch15}).

It is true that normally these instabilities become fully nonlinear and the induced motions are very complex, which are investigated mostly by simulations, but the linear stability analysis is still very useful: we can confirm that there are indeed unstable modes; the analysis of these modes and the corresponding growth rates and frequencies (if they are oscillatory) helps us unravel the mechanism of the instabilities (\citet{Guilet12}). 
The turbulence may be described as couplings of these modes.  
As a matter of fact, we conducted such linear analysis based on the Fourier (\citet{Yamasaki07}) or Laplace (\citet{Taka16}, paper I hereafter) transform for the steady and spherically symmetric post-shock accretion flows, imposing the inner and outer boundary conditions at the neutrino sphere and the standing shock wave, respectively. Although these instabilities are intrinsic, i.e., they grow from an initial perturbation on their own without any further external support, possible interactions with external perturbations are attracting much interest these days.

In fact, it has been recognized over the years (\citet{Arnett11, Muller17}) that the Si- and O-burning shells are violently convective and their turbulent motions have non-negligible influences on the shock revival when they hit the stalled shock wave.
From the point of view of the linear analysis mentioned above, this may correspond to imposing time-dependent outer boundary condition. 
Using the Laplace transform in time, paper I investigated the generation of various modes, particularly unstable ones, by the temporal fluctuations given at the outer boundary by the turbulent accreting matter. 
They demonstrated that some modes are induced more strongly than others. In this study, the inner boundary condition was left unchanged although we know that it should be also oscillatory. 
In fact, the PNS is not completely static but oscillating and wobbling owing to the exertion of impulsive forces by the matter accreting turbulently onto PNS and generates acoustic waves. 

Even if the amplitudes of the acoustic waves are not so large as to produce secondary shock waves, they may still play an important role at the linear level. 
As a matter of fact, the inner boundary condition becomes time-dependent as already mentioned and, as a result, the linearly unstable modes are expected to be affected by their presence.
This effect is what we study in this paper in the context of linear stability analysis. We use the same method as in paper I, employing the inner boundary condition that varies sinusoidally in time. We investigate how the oscillation frequencies and growth rates of various unstable modes are changed by this modification of the inner boundary condition. 

In this paper, we take also into account perturbations of the neutrino luminosity, which should be expected if PNS is wobbling. 
In so doing we consider a possible correlation between the perturbations at the outer boundary, i.e., on the shock wave and those at the inner boundary, or on the PNS surface.
Such correlations are indeed posited as a possible cause of the so-called Lepton-number Emission Self-sustained Asymmetry (LESA) (\citet{Tamborra14, TamborraPRD14}). 
In this instability observed in their numerical simulations, they found that a non-spherical configuration of lepton number emission was sustained for a long time. 
It was argued that such stable configurations were the outcomes of the interplay between the deformation of the shock wave and the anisotropic emissions of neutrinos from the PNS surface. 
Although the LESA is likely to be a phenomenon that manifests itself at nonlinear levels, it is still interesting to see what influence, if any, the correlations between the perturbations at the inner and outer boundaries may have on time-independent, linearly unstable modes. 

This paper is organized as follows. We give basic equations and explain the methods and models in the next section. In section \ref{sec:results} we present the results and discussions of linear analysis. We summarize our investigation in section \ref{sec:summary}.

\section{Methods and Models} \label{sec:methods}
In this section, we describe concisely the method we employ in this paper for linear analysis, which is based on the Laplace transform in time of the linearized hydrodynamical equations. We assume that the background flows are steady and spherically symmetric. Spherical harmonics expansion is also utilized.
The following explanations are essentially the same as those given in paper I except for the obvious changes in the treatment of the inner boundary condition.
We include them mainly for readers' convenience.

\subsection{Basic equations} \label{basic eq2}
The basic hydrodynamics equations that we employ in this paper to describe accretion flows in the supernovae core are given as follows:
\begin{eqnarray}
&\dfrac{\partial \rho}{\partial t} +\bm{\nabla} \cdot (\rho \bm{v}) = 0,& \\
&\dfrac{\partial }{\partial t}(\rho \bm{ v}) + \bm{ \nabla} \cdot (\rho \bm{ v v} + P\bm{ I}) = -\rho \dfrac{G M_{\rm{PNS}}}{r^2}\dfrac{\bm{ r}}{r},& \\
&\dfrac{\diff \varepsilon}{\diff t} +P\dfrac{\diff }{\diff t}\left( \dfrac{1}{\rho} \right) = q,&\\
&\dfrac{\partial }{\partial t}(n Y_e) +\bm{ \nabla}\cdot (nY_e \bm{ v}) = \lambda,& 
\end{eqnarray}
in addition to the equation of state (EoS), for which we adopt Shen EoS (\citet{Shen11}) in this paper. In the above equations, $\rho $, $P$, $n$, $Y_e$, $\varepsilon $ and $\bm{ v}$ are the density, pressure, number density, electron fraction, specific internal energy and velocity, respectively;  
$M_{\rm{PNS}}$ is the mass of PNS, which is assumed to be constant, and $G$ is the gravitational constant; the self-gravity of accreting matter is neglected;
we incorporate only the reactions of the electron-type neutrinos and anti-neutrinos, which are symbolically denoted by $q$ and $\lambda$ and are given by \citet{Bruenn85}.

The neutrino transfer calculation is replaced with the light-bulb approximation (\citet{Ohnishi06, Scheck06}): the luminosity is then constant in radius and is approximated in this paper as
\begin{equation}
L_{\nu_e} = \dfrac{7}{16}4\pi r_{\nu_e}^2 \sigma T_{\nu_e}^4 , \label{eq:lumi}
\end{equation}
where $\sigma $ is the Stefan-Boltzmann constant and $r_{\nu_e}$ and $T_{\nu_e}$ are the radius and temperature of the neutrino sphere for $\nu_e$; $L_{\bar{\nu}_e}$ is treated in the same way.
These neutrino luminosities ($L_{\nu_e}$ and $L_{\bar{\nu}_e}$) and temperatures ($T_{{\nu}_e}$ and $T_{\bar{{\nu}}_e}$) are model parameters. 
The radii of the neutrino sphere are determined from these parameters.

The unperturbed background flows are given as spherically symmetric steady solutions for appropriate boundary conditions. They satisfy the following equations:
\begin{eqnarray}
&\dfrac{1}{r^2} \dfrac{\diff}{\diff r}\left(\rho_0 v_{r0} r^2 \right) = 0,& \\ 
&v_{r0} \dfrac{\diff v_{r0}}{\diff r} + \dfrac{1}{\rho_0}\dfrac{\diff P_{0}}{\diff r} = - \rho_0 \dfrac{G M_{\mathrm{PNS}}}{r^2},& \\
&v_{r0} \dfrac{\diff \varepsilon_0}{\diff r} - \dfrac{P_0 v_{r0}}{{\rho_0}^2} \dfrac{\diff 
\rho_0}{\diff r} = q_0,& \\
&\rho_0 v_{r0} \dfrac{\diff Y_{e0}}{\diff r} = \lambda_0 m_b,&
\end{eqnarray}
where $m_b$ is nucleonic mass and the subscript $0$ means unperturbed quantities.
At the shock front, which is assumed to be at rest in the background flow, the Rankine-Hugoniot relations should be satisfied:
\begin{eqnarray}
 &\rho_0^{(u)} v_0^{(u)} = \rho_0^{(d)} v_0^{(d)} \\
 &\rho_0^{(u)} {v_0^{(u)}}^2 + P_0^{(u)} = \rho_0^{(d)} {v_0^{(d)}}^2 + P_0^{(d)}  \\
 & \epsilon_0^{(u)} + \dfrac{1}{2}  {v_0^{(u)}}^2 + \dfrac{P_0^{(u)}}{\rho_0^{(u)}} = \epsilon_0^{(d)} + \dfrac{1}{2}  {v_0^{(d)}}^2 + \dfrac{P_0^{(d)}}{\rho_0^{(d)}} 
\label{eq:RHeq}
\end{eqnarray}
where the superscripts (u) and (d) mean variables in the upstream and downstream of the shock, respectively.
We assume further that matter is freely falling with the radial velocity \replaced{$v_r = \sqrt{2 G M_{\rm{PNS}}/ m_b r}$}{$v_r = \sqrt{2 G M_{\rm{PNS}}/ r}$} outside the shock wave with the pressure being negligible.
By solving these equations, radius of the stationary shock wave $r_{\mathrm{sh},0}$ is determined by imposing the inner boundary condition that the density should be $10^{11} \ \mathrm{g {cm}^{-3}}$ at $r_{\nu_e}$, an approximation to the real condition that the optical depth to $r_{\nu_e}$ from infinity should be $2/3$.

Following \citet{LG00} and paper I, we expand perturbed quantities as
\begin{eqnarray}
\delta X (\bm{ r},t) &=& \sum _{l,m} \delta X^{(l, m)} (r,t) Y_{lm}(\theta , \phi)
\end{eqnarray}
where $X$ denotes scalar variables and $Y_{lm}(\theta, \phi)$ is the spherical harmonics with the indices $l$ and $m$. The velocity perturbation, on the other hand, is expanded with the vector spherical harmonics as follows:  
\begin{eqnarray}
\delta \bm{ v} (\bm{ r},t) &=& \sum _{l,m} \delta v_r^{(l, m)} (r,t) \Y \hat{\bm{ r}} \nonumber \\
 &&+\delta v _\perp ^{(l, m)} (r,t) \left[\hat{\bm{ \theta}} \dfrac{\partial Y_{lm}}{\partial \theta} + \dfrac{\hat{\bm{ \phi}}}{\sin \theta} \dfrac{\partial Y_{lm}}{\partial \phi} \right] \nonumber \\
&&+\delta v_{rot} ^{(l, m)} (r,t)\left[ -\hat{\bm{ \phi}} \dfrac{\partial Y_{lm}}{\partial \theta} + \dfrac{\hat{\bm{ \theta}}}{\sin \theta} \dfrac{\partial Y_{lm}}{\partial \phi} \right], \notag \\
\end{eqnarray}
in which the unit vectors in the spherical coordinates are denoted by $\hat{\bm{r}}, \hat{\bm{ \theta}}$, and $\hat{\bm{\phi}}$.
The linearized equations with different $(l, m)$ are decoupled from each other, since the background flow is spherically symmetric, and are written symbolically as
\begin{equation}
\label{eq.linearized}
\dfrac{\partial \bm{y}^{(l,m)}}{\partial r}(r,t) = A(r) \dfrac{\partial \bm{y}^{(l,m)}}{\partial t}(r,t) + B^{(l)}(r) \bm{y}^{(l,m)}(r,t) ,
\end{equation}
where $\bm{y}^{(l,m)}(r,t)$ denotes the vector consisting of the $(l,m)$ component of the perturbed quantities given as
\begin{equation}
\label{eq:perturbedstate}
\bm{ y}(r,t) = \left(\dfrac{\delta \rho}{\rho _0}, \dfrac{\delta v_r}{v_{r0} }, \dfrac{\delta v_\perp}{v_{r0} }, \dfrac{\delta \varepsilon}{\varepsilon_0}, \dfrac{\delta Y_e}{Y_{e0} }, \dfrac{\delta v_{rot}}{v_{r0} } \right)^T,
\end{equation}
where $(\cdots)^T$ means transposition. 
Note that we take $\rho$, $\varepsilon$, $Y_{e}$ as independent thermodynamic variables. 
The coefficient matrices, $A(r)$ and $B^{(l)}(r)$, are made of the unperturbed quantities alone (see appendix A in paper I) and are independent of $m$ because of the spherical symmetry of the background. 
We solved the linearized equations (\ref{eq.linearized}) in the region between the standing shock ($r=r_{\mathrm{sh,0}}$) and the PNS surface ($r=r_{\nu_e}$) in the unperturbed state. Hereafter the subscripts $0$ and $(l,m)$ are omitted for notational simplicity.

The outer boundary condition imposed at the shock radius is given by the linearized Rankine-Hugoniot relations, which are schematically written as
\begin{equation}
\label{eq.outerboundary}
\bm{ y}(r_\mathrm{sh},t) = R\bm{ z}(t) +\dfrac{\partial }{\partial t}\dfrac{\delta r_\mathrm{sh}(t)}{r_\mathrm{sh}} \bm{c} +\dfrac{\delta r_\mathrm{sh}(t)}{r_\mathrm{sh}} \bm{d},
\end{equation}
where $\delta r_\mathrm{sh}(t)$ is the time-dependent perturbation to the shock radius; $R$ is a matrix and $\bm{c}$ and $\bm{d}$ are vectors, which depend only on the background quantities, whereas $\bm{z}$ is the perturbation in the upstream flow (see also appendix A in paper I), which may be originated from turbulent convection in the envelope. 

The inner boundary is set at the PNS surface. Since the perturbation of the shock radius is the only variable remaining after imposing the outer boundary condition, we can give only one condition at the inner boundary. It is symbolically written as
\begin{equation}
\label{eq.innerboundary}
f(\bm{y}(r_{\nu _e},t),t) = 0.
\end{equation}

In principle, we can set any initial condition to the perturbation:
\begin{equation}
\label{eq.initial}
\bm{y}(r,t=0) = \bm{ y}_0(r) \ \ (r_{\nu _e} < r < r_\mathrm{sh}).
\end{equation}
In this paper, however, we set $\bm{y}_0(r) = 0$ for simplicity. We are hence concerned only with the perturbations generated at the boundaries.

To summarize, the problem is now reduced to solving equations~(\ref{eq.linearized}), (\ref{eq.outerboundary}), (\ref{eq.innerboundary}) and (\ref{eq.initial}) for the perturbation to the shock radius, $\delta r_\mathrm{sh}/r_\mathrm{sh}(t)$, as an initial-boundary-value problem.

\begin{deluxetable*}{lcccccc}[htbp!]
 \tablecaption{Comparison of three models in this paper \label{tab:models}}
 \tablehead{
  & \colhead{\shortstack[l]{Acoustic \\ injection}}  & \colhead{\shortstack[l]{Perturbations of \\ neutrino luminosity}} & \colhead{Basic equation} & \colhead{\shortstack[l]{Outer boundary \\condition}} & \colhead{\shortstack[l]{\\Degree of freedom(s) \\after imposing \\outer boundary condition}} & \colhead{\shortstack[l]{Inner boundary\\ condition(s)}}
}
 \startdata
 Model A & no  & no  & (\ref{eq.L-linearized}) & (\ref{OB}) & $\delta r_\mathrm{sh}^*/r_\mathrm{sh}$ & (\ref{eq:IBfixed}) \\
 Model B & yes & no  & (\ref{eq.L-linearized}) & (\ref{OB}) & $\delta r_\mathrm{sh}^*/r_\mathrm{sh}$ & (\ref{eq.outgoing}) \\
 Model C & yes & yes & (\ref{eq.leqLESA}) & (\ref{OB}) & $\delta r_\mathrm{sh}^*/r_\mathrm{sh}$ and $\delta T_{\alpha}^*/T_{\alpha}$ & (\ref{eq.outgoing}) and (\ref{eq.IBLESA}) \\
 \enddata
\end{deluxetable*}

\subsection{Laplace transformation of linearized system} \label{sec.modes}
To solve this initial-boundary-value problem, we use the Laplace transform with respect to time defined as 
\begin{equation}
 f^{*}(s) := \int_0^{\infty} f(t) \e^{-st} \diff t 
\end{equation}
where $s$ is a complex variable. Hereafter, the superscript $*$ means Laplace-transformed functions, which are complex in general.
Equations (\ref{eq.linearized}), (\ref{eq.outerboundary}) and (\ref{eq.innerboundary}) are Laplace-transformed into the following forms:
\begin{eqnarray}
\label{eq.L-linearized}
&\dfrac{\diff \bm{ y}^* }{\diff r} (r,s) = (sA+B)\bm{ y}^*(r,s) -A\bm{ y}_0(r),& \label{ODE} \\
\label{OB}
&\bm{ y}^*(r_\mathrm{sh},s) = (s\bm{ c} + \bm{ d})\dfrac{\delta r_\mathrm{sh}^* (s)}{r_\mathrm{sh}} +R\bm{ z}^*(s),& \\
\label{IB}
&f^*(\bm{ y}^*(r_{\nu _e},s),s) = 0.&
\end{eqnarray}
In linear analysis, the inner boundary condition is generally written as
\begin{equation}
 f^*(\bm{ y}^*(r_{\nu _e},s),s) = \bm{ a }^*(s)\cdot \bm{ y}^* (r_{\nu _e},s) + b^*(s)= 0, \label{eq.IB}
\end{equation}
where $\bm{ a}^*$ and $b^*$ are some functions of $s$. 
 Equation (\ref{ODE}) is a system of ordinary differential equations and can be easily integrated.
Combined with equations (\ref{OB}) and (\ref{IB}), they determine $\delta r_\mathrm{sh}^*/r_\mathrm{sh}(s)$.

Following the common practice, we assume that $\delta r_\mathrm{sh}/r_\mathrm{sh}(t)$ is written as a superposition of eigenmodes as 
\begin{equation}
 \dfrac{\delta r_\mathrm{sh}}{r_\mathrm{sh}}(t)  = \sum _{j} a_j \e^{\Omega_j t} \e^{i (\omega_j t + \phi_j)} \label{eq.superposition},
\end{equation}
where $\Omega _j$ and $\omega _j$ are the growth or damping rate and the oscillation frequency of the $j$-th mode ($j = 1, 2,3,\cdots $), respectively,  and $a_j$ is the amplitude of the same mode, which is independent of $t$.
 We can assume $\omega _j \ge 0$ and $-\pi/2 \le \phi _j < \pi /2$ for all $j$ without loss of generality.
Then the Laplace transformation of $\delta r_\mathrm{sh}/r_\mathrm{sh}(t)$ is written as 
\begin{eqnarray}
 \label{shock_evolution}
 \dfrac{\delta r_\mathrm{sh}^*}{r_\mathrm{sh}}(s)  = \sum _j a_j\dfrac{\e^{i \phi_j}}{(s-\Omega_j) - i \omega _j},
\end{eqnarray}
which has poles at $\Omega_j + i \omega_j$  ($j = 1, 2,3,\cdots $). 
The stability or instability of the standing shock can be judged from the sign of $\Omega_j$.

\subsection{Model parameters and the treatment of neutrino heating and cooling in the unperturbed flows} 
We assume  $L_\nu := L_{\nu _e}= L _{\bar{\nu}_e}$ for simplicity and change its value as a free parameter. 
The values of other parameters that specify the unperturbed background flow are set as follows: the mass of PNS is $M_{\mathrm{PNS}} = 1.4 M_\odot$;
the mass accretion rate and neutrino temperatures are fixed to $\dot{M} = 0.6$ M$_\odot$  s$^{-1}$ and $T_{\nu_e} = T_{\bar{\nu}_e} = 4.5$ MeV, respectively;
the entropy and $Y_e$ just ahead of the shock wave are set as $S = 3k_B$ and $Y_e = 0.5$, respectively, where $k_B$ is the Boltzmann constant;
matter is assumed to free-fall from infinity onto the shock.
We employ Shen's EoS \citep{Shen11}, which takes into account the contributions from nucleons, nuclei, $\alpha$ particles, photons, electrons and positrons. 

The neutrino heating and cooling functions, $q$ and $\lambda$, are evaluated under the light bulb approximation as follows: 
\begin{eqnarray}
 q = - && \sum_{\alpha} \dfrac{1}{\rho} \dfrac{4 \pi c}{\left(2 \pi \hbar c\right)^3} \times  \notag \\ 
&& \int_0^{\infty} \diff \epsilon \, \epsilon^3 \left[j_{\alpha }(\epsilon) - \left(j_{\alpha }(\epsilon) + \kappa_{\alpha}(\epsilon) \right) f_{\alpha}(\bm{ x}, \epsilon)\right], \ \  \label{eq:q}\\
 \lambda = - && \sum_{\alpha} i_{\alpha } \dfrac{m_b}{\rho} \dfrac{4 \pi c}{\left(2 \pi \hbar c\right)^3} \times \notag \\ 
&&\int_0^{\infty} \diff \epsilon \, \epsilon^2 \left[j_{\alpha }(\epsilon) - \left(j_{\alpha }(\epsilon) + \kappa_{\alpha}(\epsilon) \right) f_{\alpha}(\bm{ x}, \epsilon)\right], \ \ \label{eq:lambda}
\end{eqnarray}
where $\alpha$ specifies the neutrino species and $i_{\alpha}$ is defined as
\begin{equation}
 i_{\alpha} =
  \begin{cases}
   1  & \text{for} \  \nu_e \\
   -1 & \text{for} \ \bar{\nu}_e.
  \end{cases}
\end{equation}
 $\epsilon$ denotes the neutrino energy and $\kappa_{\alpha}$ and $j_{\alpha}$ are the absorptivity and emissivity of each neutrino species, for which we employed the Bruenn's rates (\citet{Bruenn85}).
The distribution function of neutrinos is denoted by $f_{\alpha}$ and is approximated by the scaled Fermi-Dirac distribution with a vanishing chemical potential:
\begin{equation}
 f_{\alpha}(\bm{ x}, \epsilon) = \dfrac{1}{1 + \exp(\epsilon/k_{B} T_{\alpha})} g(r),
\end{equation} 
where $g(r)$ is the so-called geometrical factor defined as
\begin{equation}
 g(r) = \dfrac{1 - \sqrt{ 1 - \left( r_{\nu}/r \right)^2}}{2},
\end{equation}
as a function of $r = |\bm{ x}|$. See \citet{Ohnishi06} for more details.

The unperturbed flow models employed in this paper are the same as those adopted in paper I. The radius of neutrino sphere, the characteristic frequencies of advective-acoustic and purely acoustic cycles and the gain radius for these background models are listed in Table 1 of paper I. 

\subsection{Injection of acoustic waves and perturbations of neutrino luminosity}\label{sec.model}
As new ingredients in this paper, we analyze effects of the injection of acoustic waves from the inner boundary as well as of the fluctuations of the neutrino luminosity.
We introduce the former as a time-dependent inner boundary condition. In considering the latter, on the other hand, we introduce the fluctuation of neutrino temperature as a new degree of freedom and impose an additional inner boundary condition.

In the following, we give the details of the numerical treatments of these two ingredients in turn. Table \ref{tab:models} is a concise summary of the three models considered in this paper. 

\subsubsection{Injection of acoustic waves}
According to the general solution of the linearized equations (see equations (\ref{eq:GeneralSol}) and (\ref{eq:det}) in appendix \ref{sec:formal sol.}), the positions of poles in the complex plane are affected directly by the inner boundary condition through the coefficient $\bm{a}^*(s)$ in equation (\ref{eq.IB}). 
This is in sharp contrast to the outer boundary condition, which has only an indirect leverage.
It is hence important to give an appropriate condition at the inner boundary.

It should be noted that the acoustic mode has been already taken into account in the linearized equations.
This is understood as follows.
The propagation speeds of eigenmodes are the eigenvalues of the matrix $V$ in the linearized equations written as
\begin{equation}
 \dfrac{\partial \bm{y}}{\partial t} + V \dfrac{\partial \bm{y}}{\partial r} + A^{-1} B \bm{y} = 0.
\end{equation}
They are actually $v_r$, $v_r - c_s$, $v_r + c_s$, where $c_s$ is the sound speed. Whereas $v_r$ is quadruply degenerate with the corresponding eigenmodes being $\delta v_{\perp}$, $\delta v_{\mathrm{rot}}$, $\delta Y_e$, $P/(\rho)^2 \delta \rho - \delta \varepsilon$, $v_r - c_s$ and $v_r + c_s$ correspond, respectively, to the ingoing and outgoing acoustic modes, which have the eigenvectors expressed as
\begin{eqnarray}
 y_{\mathrm{in}} &=&\dfrac{1}{{c_s}^2} \left(\dfrac{\partial P}{\partial \rho}\right) \dfrac{\delta \rho}{\rho} - \dfrac{v_r}{c_s} \dfrac{\delta v_r}{v_r} \notag \\
&& + \dfrac{\varepsilon}{{c_s}^2 \rho} \left(\dfrac{\partial P}{\partial \varepsilon} \right) \dfrac{\delta \varepsilon}{\varepsilon} +\dfrac{Y_e}{{c_s}^2 \rho} \left(\dfrac{\partial P}{\partial Y_e} \right) \dfrac{\delta Y_e}{Y_e} \\
 &=& \dfrac{1}{{c_s}^2 \rho} \delta P - \dfrac{v_r}{c_s} \dfrac{\delta v_r}{v_r}, \\ 
 y_{\mathrm{out}} &=& \dfrac{1}{{c_s}^2 \rho} \delta P + \dfrac{v_r}{c_s} \dfrac{\delta v_r}{v_r}.
\end{eqnarray}

Based on this observation, we impose the following condition at the inner boundary to inject the acoustic waves, which may be produced by the $g$-mode oscillation of PNS:
\begin{equation}
 y_{\mathrm{out}}(r_{\nu_e}, t) = \alpha \sin \left(\omega_{\mathrm{PNS}} t\right), \label{eq.outgoing}
\end{equation}
where the amplitude $\alpha $ is a free parameter, which we set to 1. Laplace-transformed, the right hand side of the above equation gives $b^*(s)$ in equation (\ref{eq.IB}). 
As can be seen from the general solution (\ref{eq:det}) of the linearized equations, $b^*(s)$ does not affect the pole positions of $\delta r_\mathrm{sh}^*/r_\mathrm{sh}(s)$, or the stability of shock wave. As for $\omega_{\mathrm{PNS}}$ in equation (\ref{eq.outgoing}), we employ the typical g-mode frequency of PNS:
\begin{equation}
 \omega_{\mathrm{PNS}} = 2000 \times l \ \mathrm{s}^{-1}
\end{equation}
(\citet{Burrows06}).
Note in passing that in paper I, we imposed the following inner boundary condition:
\begin{equation}
 \delta v_{r} = 0. \label{eq:IBfixed}
\end{equation}

\subsubsection{Fluctuations of neutrino luminosity}
To model the perturbation of neutrino luminosity, we introduce a new degree of freedom, i.e. the fluctuation of neutrino temperature $\delta T_{\alpha}$, and expand it as usual:
\begin{equation}
 \delta T_{\alpha} = \sum _{l,m} \delta T_{\alpha}^{(l, m)} (t) Y_{lm}(\theta , \phi),
\end{equation}
which is consistent with the black body approximation employed for the neutrino luminosity (see equation (\ref{eq:lumi})).
We determine $\delta T_{\alpha}$ as follows. We assume that it is related with the perturbation to $Y_e$ in the vicinity of the neutrino sphere.
Indeed for each $(l, m)$ with $l > 1$, we impose the following relation:

\begin{equation}
 \left(\dfrac{\partial P}{\partial Y_e}\right)_{\rho, T} \delta Y_e^{(l,m)}(r_{\nu_e}, t) +  \left(\dfrac{\partial P}{\partial T}\right)_{\rho, Y_e} \delta T_{\alpha}^{(l,m)}(t) = 0. \label{eq.IBLESA}
\end{equation}
This means that $\delta T_{\alpha}$ is equal to the perturbation to the matter temperature that could cancel the pressure fluctuation that the $Y_e$ perturbation would induce (\citet{Janka16}). We further assume that there is no spherically symmetric ($l = 0$) perturbation to the neutrino temperature. 
\added{Note that although the emission of $\nu_e$ and $\bar{\nu}_e$ via electron/positron captures in the cooling region, which gives a substantial contribution to the neutrino luminosity in fact (\citet{Muller12}), is not incorporated explicitly in this paper, its effects on the shock instability are taken into account effectively through the correlation between the perturbation of neutrino luminosity and the fluctuation in the accreting matter given above, which are also expected to the neutrino emission in the cooling region.}

The fluctuation of the neutrino temperature affects $q$ and $\lambda$:
\begin{eqnarray}
 \delta q = \dfrac{\partial q}{\partial \rho} \delta \rho + \dfrac{\partial q}{\partial \varepsilon} \delta \varepsilon + \dfrac{\partial q}{\partial Y_e} \delta Y_e + \delta q_{\nu}, \\
\delta \lambda = \dfrac{\partial \lambda}{\partial \rho} \delta \rho + \dfrac{\partial \lambda}{\partial \varepsilon} \delta \varepsilon + \dfrac{\partial \lambda}{\partial Y_e} \delta Y_e + \delta \lambda_{\nu}, \label{eq.lambda_nu}
\end{eqnarray}
where $\delta q_{\nu}$ and $\delta \lambda_{\nu}$ are the new terms.
They are written as
\begin{eqnarray}
  \delta q_{\nu} =  && \sum_{\alpha} \dfrac{1}{\rho} \dfrac{4 \pi c}{\left(2 \pi \hbar c\right)^3} \times \notag \\
&& \int_0^{\infty} \diff \epsilon \, \epsilon^3 \left[\left(j_{\alpha }(\epsilon) + \kappa_{\alpha}(\epsilon) \right) \delta f_{\alpha}(\bm{ x}, \epsilon)\right], \label{eq:dqLESA}\\
 \delta \lambda_{\nu} = && \sum_{\alpha} i_{\alpha} \dfrac{m_b}{\rho} \dfrac{4 \pi c}{\left(2 \pi \hbar c\right)^3} \times \notag \\
&& \int_0^{\infty} \diff \epsilon \, \epsilon^2 \left[\left(j_{\alpha }(\epsilon) + \kappa_{\alpha}(\epsilon) \right) \delta f_{\alpha}(\bm{ x}, \epsilon)\right],
\label{eq:dlambdaLESA}
\end{eqnarray}
where the perturbation to the neutrino distribution is given as
\begin{equation}
 \delta f_{\alpha}(\bm{x}, \epsilon) = \dfrac{\beta_{\alpha} \epsilon \e^{\beta_{\alpha} \epsilon }}{\left(1 + \e^{\beta_{\alpha} \epsilon}\right)^2} \, g(|\bm{x}|) \dfrac{\delta T_{\alpha}}{T_{\alpha}}
\end{equation}
with $\beta_{\alpha} = 1/k_{B} T_{\alpha}$. 
Equation (\ref{eq.linearized}) is then modified as
\begin{equation}
 \dfrac{\partial \bm{y}}{\partial r} = A\dfrac{\partial \bm{y}}{\partial t} +B\bm{y} + \bm{u} \dfrac{\delta T_{\alpha}}{T_{\alpha}}, \label{eq.basicLESA}
\end{equation}
where $\bm{u}$ is related with $\delta q_{\nu}$ and $\delta \lambda_{\nu}$ and its explicit form is given in Appendix \ref{sec:v}.
Finally, the Laplace-transformed linearized equation is obtained as 
\begin{equation}
 \dfrac{\diff \bm{ y}^* }{\diff r} (r,s) = (sA+B)\bm{ y}^*(r,s) -A\bm{ y}_0(r) + \bm{u} \dfrac{\delta T_{\alpha}^*}{T_{\alpha}}. \label{eq.leqLESA}
\end{equation}

In summary, we solve the linear equation (\ref{eq.leqLESA}) when the perturbation to the neutrino luminosity is taken into account. Then the remaining degrees of freedom are $\delta r_{\mathrm{sh}}^*$ and $\delta T_{\alpha}^*$ after imposing the Rankine-Hugoniot relations at the outer boundary, which are determined from the two conditions given in equations (\ref{eq.IB}) and (\ref{eq.IBLESA}). The general form of the solution is given in Appendix \ref{sec:formal sol.}.

\begin{figure*}[htbp!]
  \gridline{
 \fig{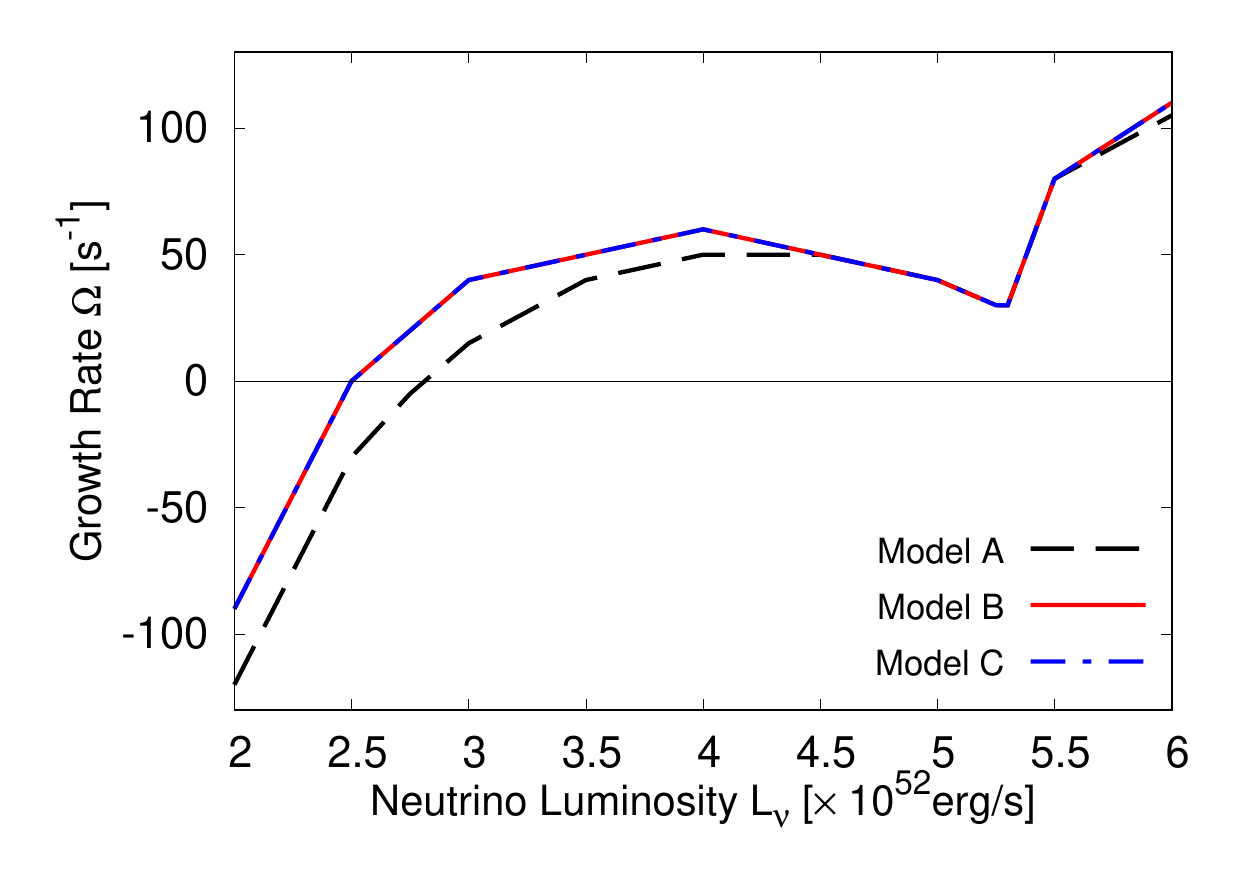}{0.5 \textwidth}{(a) Growth rates ($l = 1$)}
 \fig{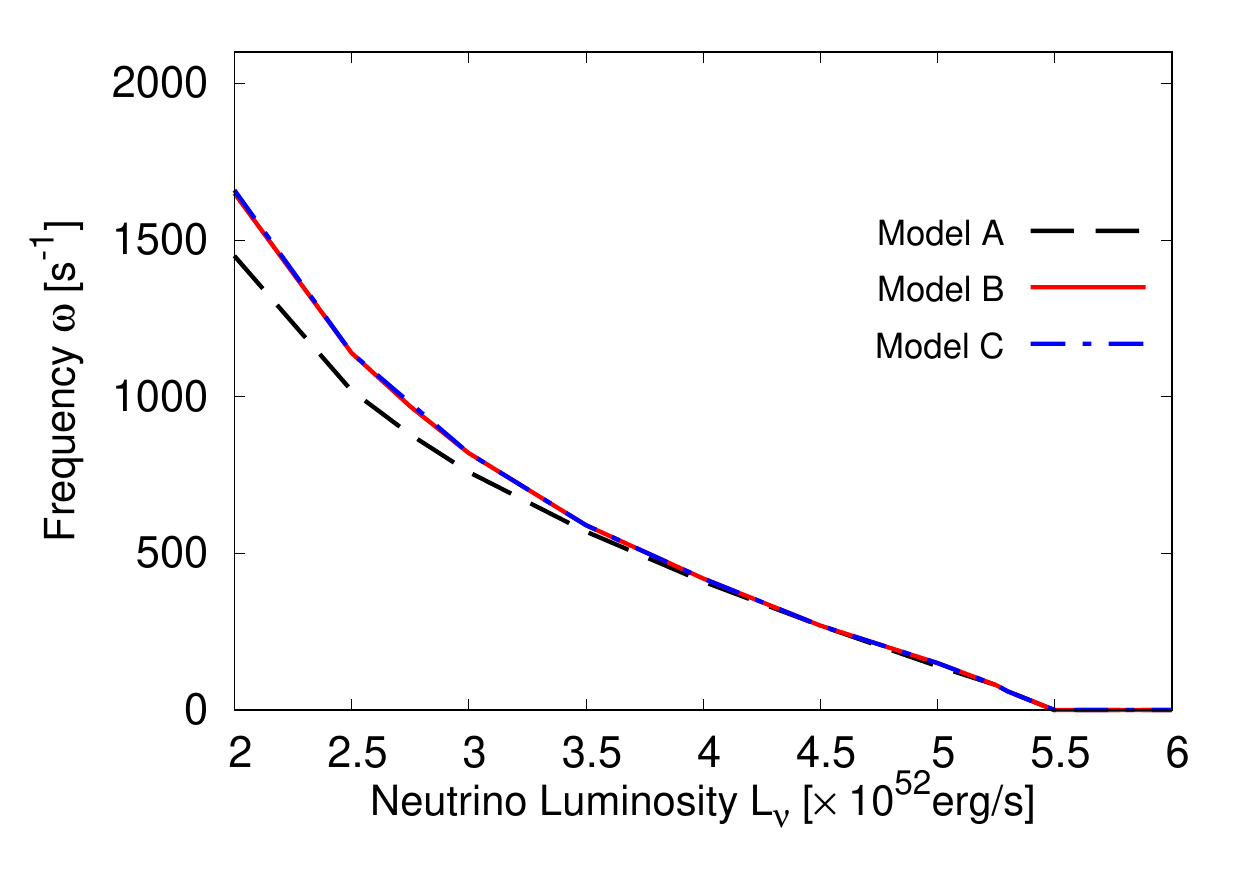}{0.5 \textwidth}{(b) Frequencies ($l = 1$)}}
  \caption{The growth rates (a) and oscillation frequencies (b) of the dominant eigenmodes for three models as a function of the neutrino luminosity $L_{\nu}$. Black dashed, red solid and blue dash-dotted lines show the results of models A, B and C, respectively.}
\label{fig:pole12}
 \end{figure*}

\section{Results and discussions} \label{sec:results}
\subsection{Instabilities of the standing shock}
\begin{figure*}[htbp]
 \gridline{
 \fig{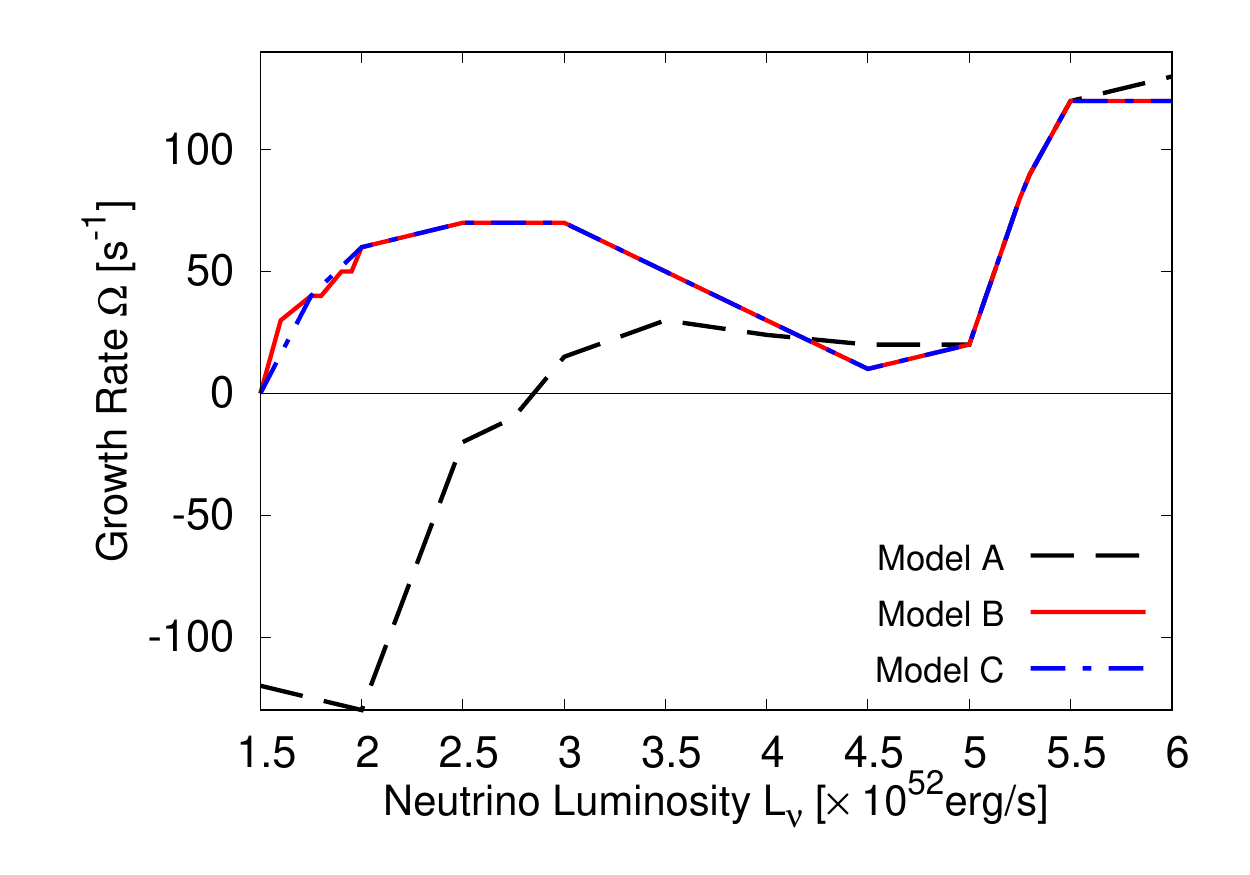}{0.5 \textwidth}{(a) Growth rates ($l = 2$)}
 \fig{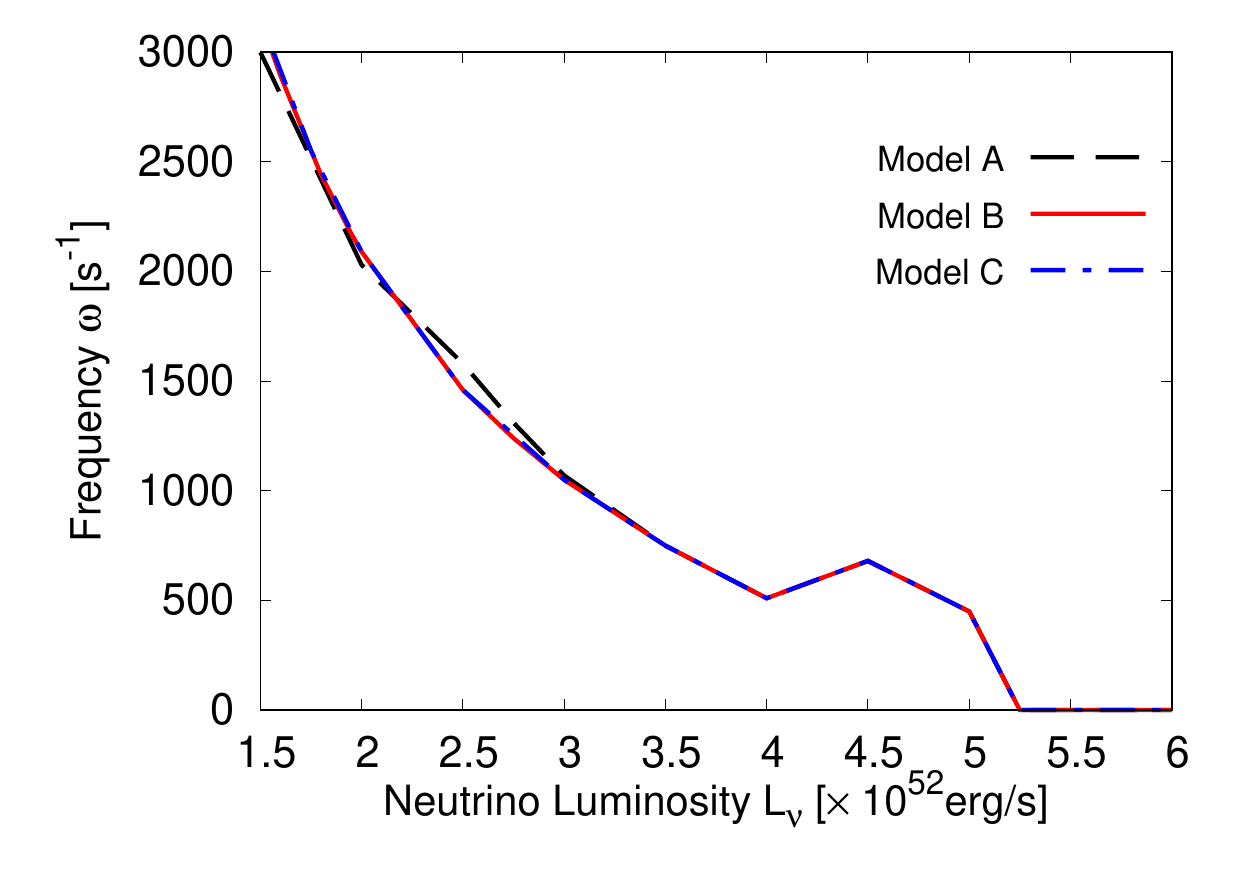}{0.5 \textwidth}{(b) Frequencies ($l = 2$)}}
  \gridline{
 \fig{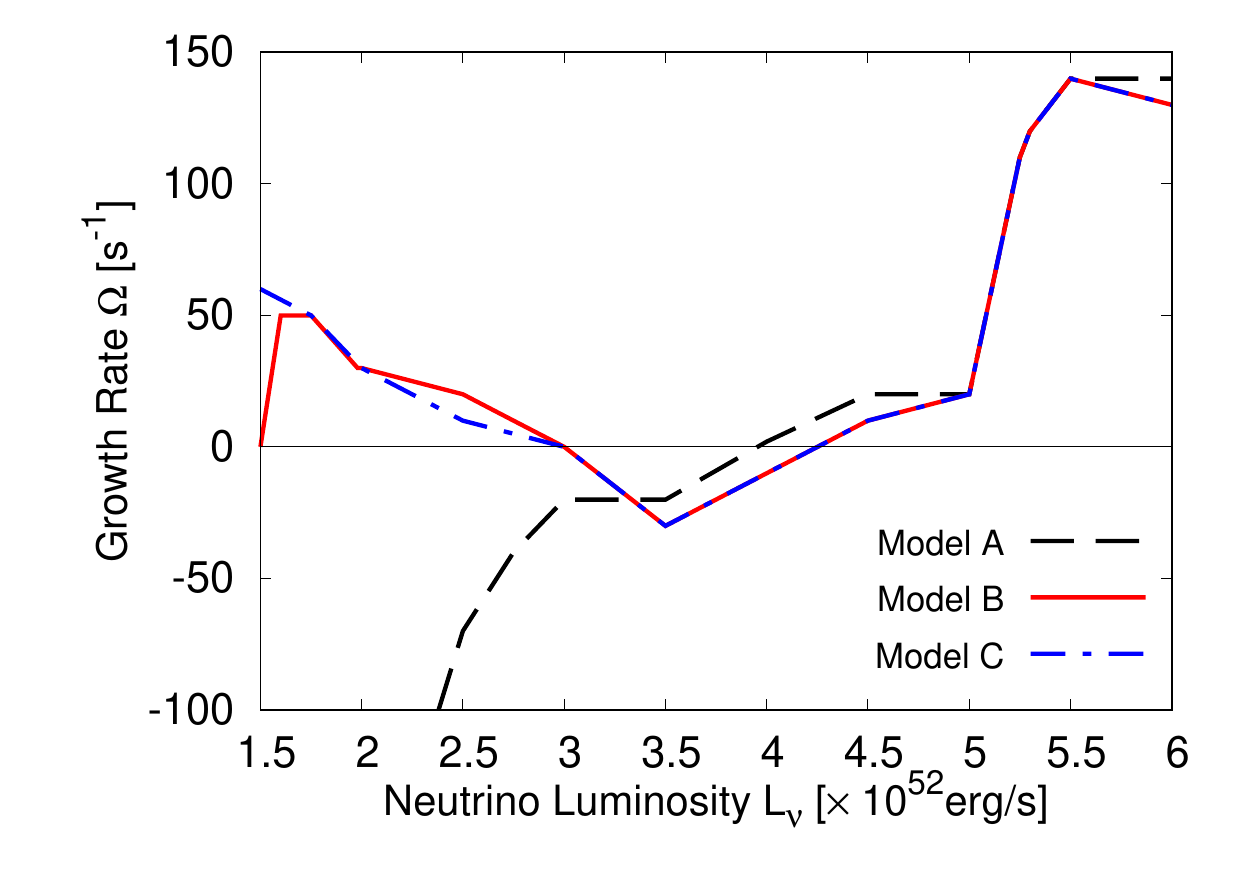}{0.5 \textwidth}{(c) Growth rates ($l = 3$)}
 \fig{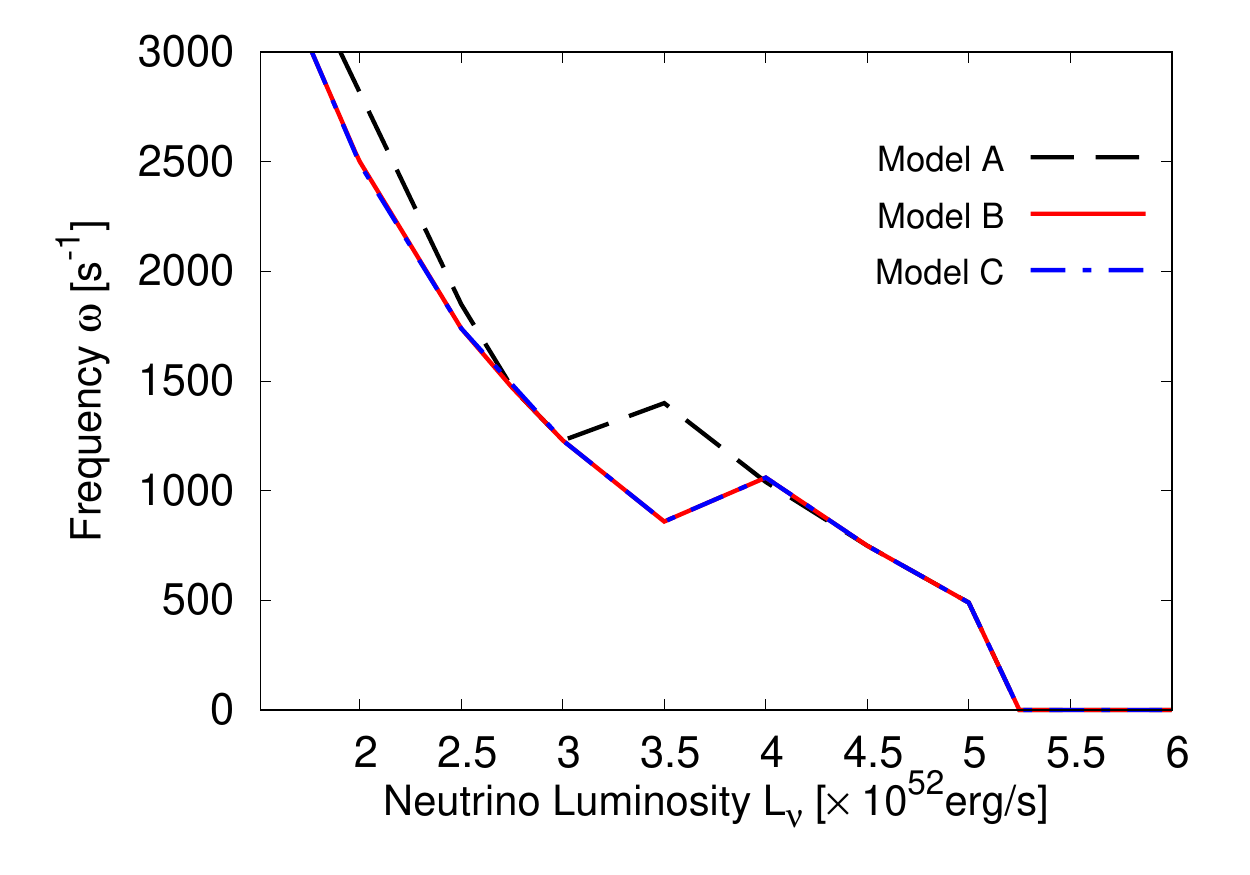}{0.5 \textwidth}{(d) Frequencies ($l = 3$)}}
  \gridline{
 \fig{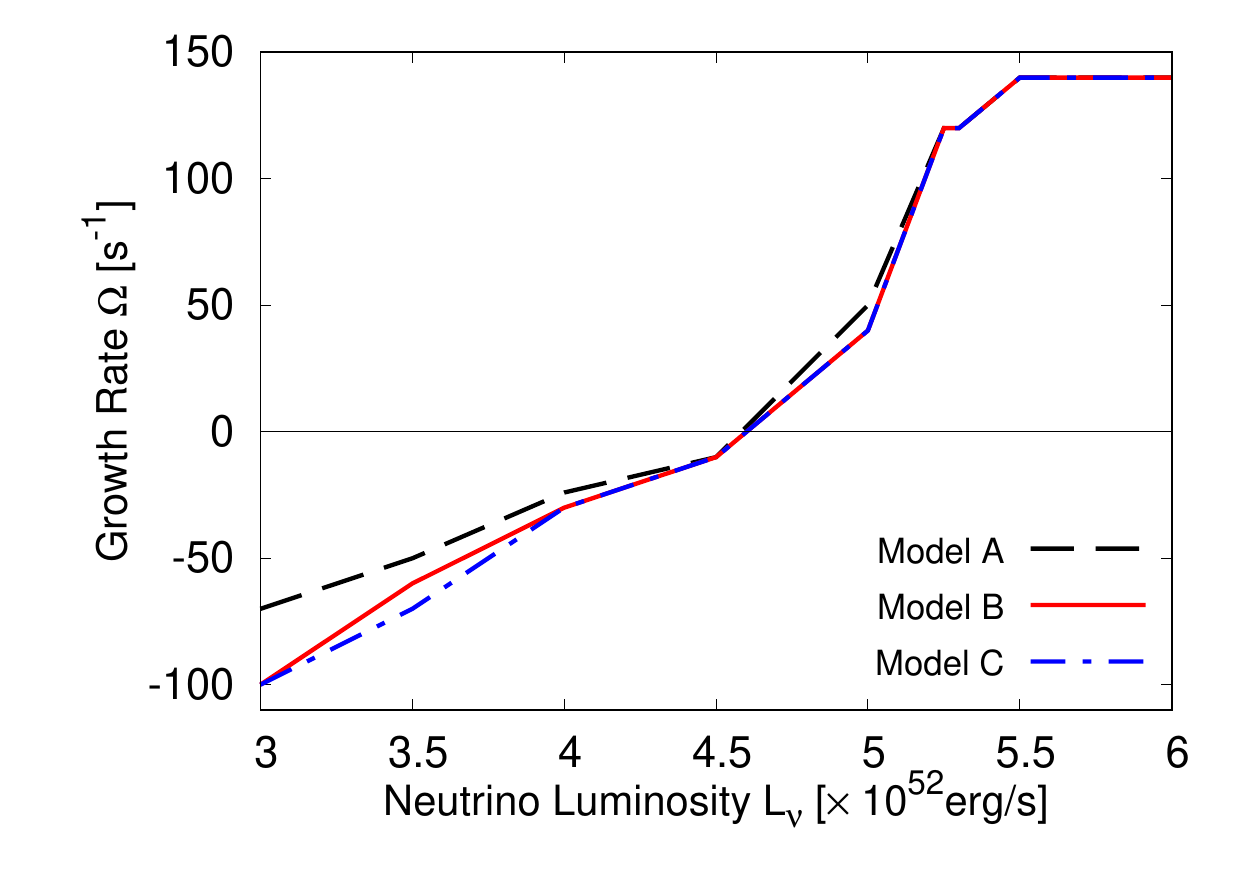}{0.5 \textwidth}{(e) Growth rates ($l = 4$)}
 \fig{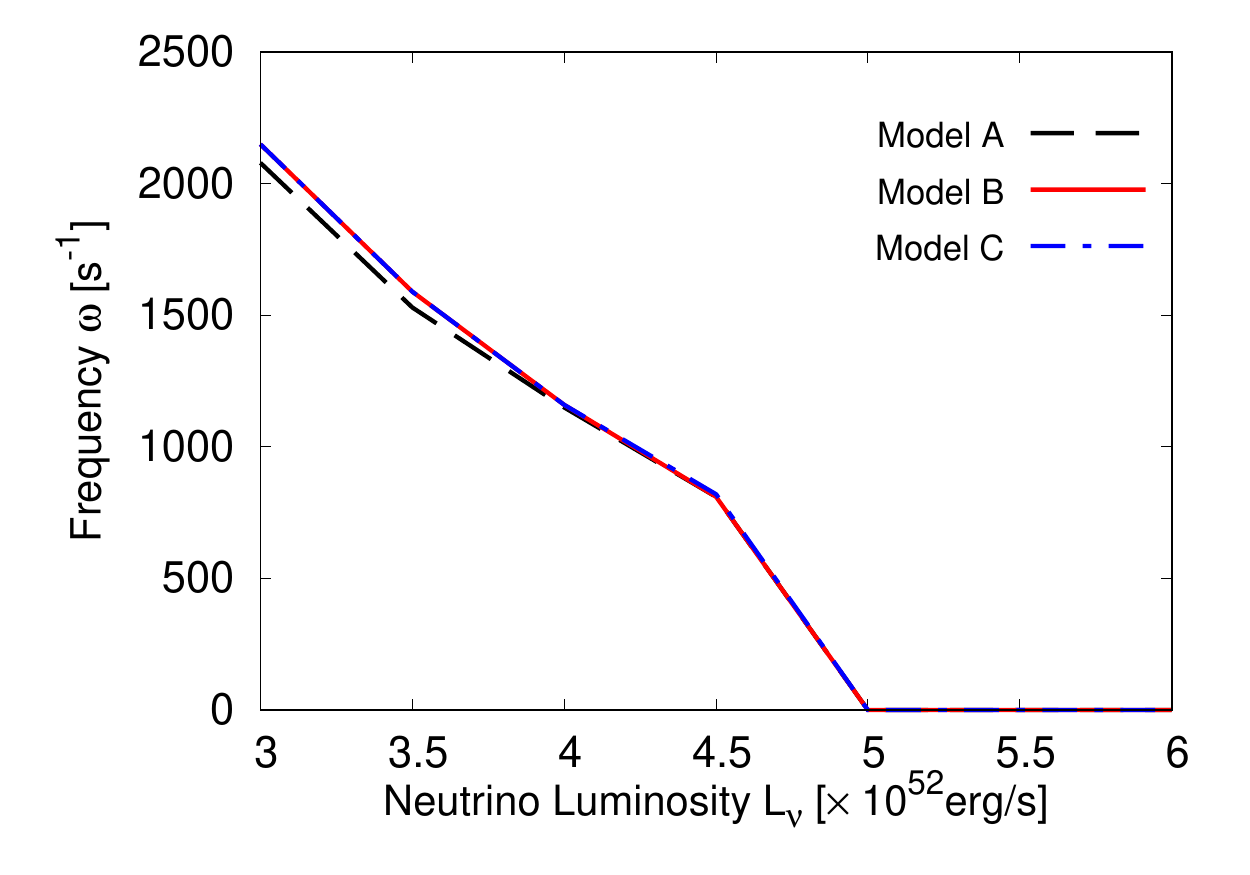}{0.5 \textwidth}{(f) Frequencies ($l = 4$)}}
  \caption{Same as Figure \protect\ref{fig:pole12} but for higher $l$ modes.}
\label{fig:pole34}
\end{figure*}

We first present the growth rates as well as the oscillation frequencies of the dominant modes for models A, B and C as a function of the neutrino luminosity. They are read out from the positions of the corresponding poles of $\delta r_{\mathrm{sh}}^*/r_{\mathrm{sh}}(s)$.
Note that there are several poles for a given luminosity in fact (see Figures 3 and 4 in paper I). 
We plot the results for $l = 1$ in Figure \ref{fig:pole12}. The black, red and blue lines correspond to models A, B and C, respectively.
Note again that model A is the same as that studied in paper I.
Figure \ref{fig:pole34} gives the results for higher $l$ modes.

We first look at the effect of the acoustic injection. From a comparison of the different models in Figure \ref{fig:pole12} (a), the growth rate of the dominant mode with $l = 1$ is enhanced by the acoustic power injection especially when the neutrino luminosity is low $L_{\nu} \ltsim 4.0 \times 10^{52} \ \mathrm{erg \  s^{-1}}$. 
This enhancement is accompanied by the lowering of the critical luminosity from $L_{\nu} \sim 2.8 \times 10^{52} \ \mathrm{erg \  s^{-1}}$ to $L_{\nu} \sim 2.5 \times 10^{52} \ \mathrm{erg \  s^{-1}}$.
By the critical luminosity we mean here the lowest luminosity, at which this particular mode becomes unstable.
At high neutrino luminosities $ L_{\nu} \gtsim 4.5 \times 10^{52} \ \mathrm{erg \  s^{-1}}$ the acoustic power does not affect the growth rates very much.
Note that the oscillation frequency vanishes at $L_{\nu} \gtsim 5.5 \times 10^{52} \ \mathrm{erg \  s^{-1}}$. This may be interpreted as the transition from the SASI regime to the convection regime. 
Then we may say that the acoustic injection does not play an important role in the convective instability in the linear order.

It is not the $l = 1$ mode but $l = 2$ and $3$ mode that have the greatest growth rate at low neutrino luminosities.  
As we can see from Figure \ref{fig:pole34}~(a) and~(c), this is due to the strong enhancements of the growth rates at $L_{\nu} \ltsim 4.0 \times 10^{52} \ \mathrm{erg \  s^{-1}}$.
The critical luminosities for these modes are also significantly reduced.
These results indicate that quadrupole and octupole radiations of acoustic wave from PNS may strongly enhance the instability of the same anisotropies.

For $l \geq 4$ modes, the modifications of the inner boundary condition do not have much influence on the shock instability. As an example, the $l = 4$ case is shown in panels (e) and (f) of the same figure. The growth rates of the three eigenmodes decrease monotonically in a similar way as the neutrino luminosity gets smaller and the critical luminosity is not changed much either. This situation is common to other high $l$ modes.

 Compared with the growth rates, the oscillation frequencies are less affected by the modifications of the inner boundary conditions as is evident from the right panels in Figures \ref{fig:pole12} and \ref{fig:pole34}. This may be because the eigenfrequency of the shock is simply determined by the global structure of the background flow below the shock wave. 

There appear humps around $L_{\nu} \sim 4.0 - 4.5 \times 10^{52} \ \mathrm{erg \  s^{-1}}$ in panels (b) and (d) in Figure \ref{fig:pole34}. This happens because the second overtones take over the position of the maximum-growth-rate mode (see Figure 3 in paper I for such behavior).

Finally the comparison of models B and C in Figures \ref{fig:pole12} and \ref{fig:pole34} shows that the perturbation of neutrino luminosity changes neither the growth rates nor the oscillation frequencies appreciably. This is common to all the modes studied here and suggests that the fluctuations of neutrino luminosity may not be important for the shock instability in the linear order.

\subsection{Instigations of various modes}
Various modes, stable and unstable alike, are excited by the perturbations propagated from the inner boundary, at which we impose the time-dependent conditions. We hence need not set a non-vanishing perturbation initially. How strongly they are instigated is different from mode to mode and given by the amplitude $a_j$ of equation (\ref{eq.superposition}), which can be obtained by calculating the residue of $\delta r_{\mathrm{sh}}^*/r_{\mathrm{sh}}(s)$ (see appendices D, E and F in paper I for details).
Figure \ref{fig:amplitude} shows the absolute values of the amplitudes for the dominant unstable eigenmodes with $l = 1, 2, 3$ and $4$ in the spherical harmonics expansion. 
 Dash-dotted lines and squares correspond to model A, solid lines and crosses mean model B and dashes lines and circles represent model C. 
Colors indicate the different neutrino luminosities. 

Comparing the results for models A and B, one finds that the injection of the acoustic power magnifies the amplitudes in general and that the amplification is remarkable when the neutrino luminosity is low. 
In fact, the enhancement is as high as a few orders of magnitude at low luminosities whereas it is just a factor of a few at high luminosities. It is also evident that the enhancement is more remarkable for $l = 1$ and 2 modes. 
The suppression of the enhancement at high luminosities is due to the mismatch between the frequencies of g-modes in PNS and those of SASI and convection. 
This will be explained more in detail later. 

The comparison of models B and C in Figure \ref{fig:amplitude} shows, on the other hand, that the excitation of modes is little affected by the perturbation of neutrino luminosity. 
This is consistent with the previous findings on the growth rate and oscillation frequency and suggests that the temporal fluctuation of the neutrino luminosity is not very important for the shock instability at least in the linear order.

\begin{figure*}[tbhp]
  \gridline{
 \fig{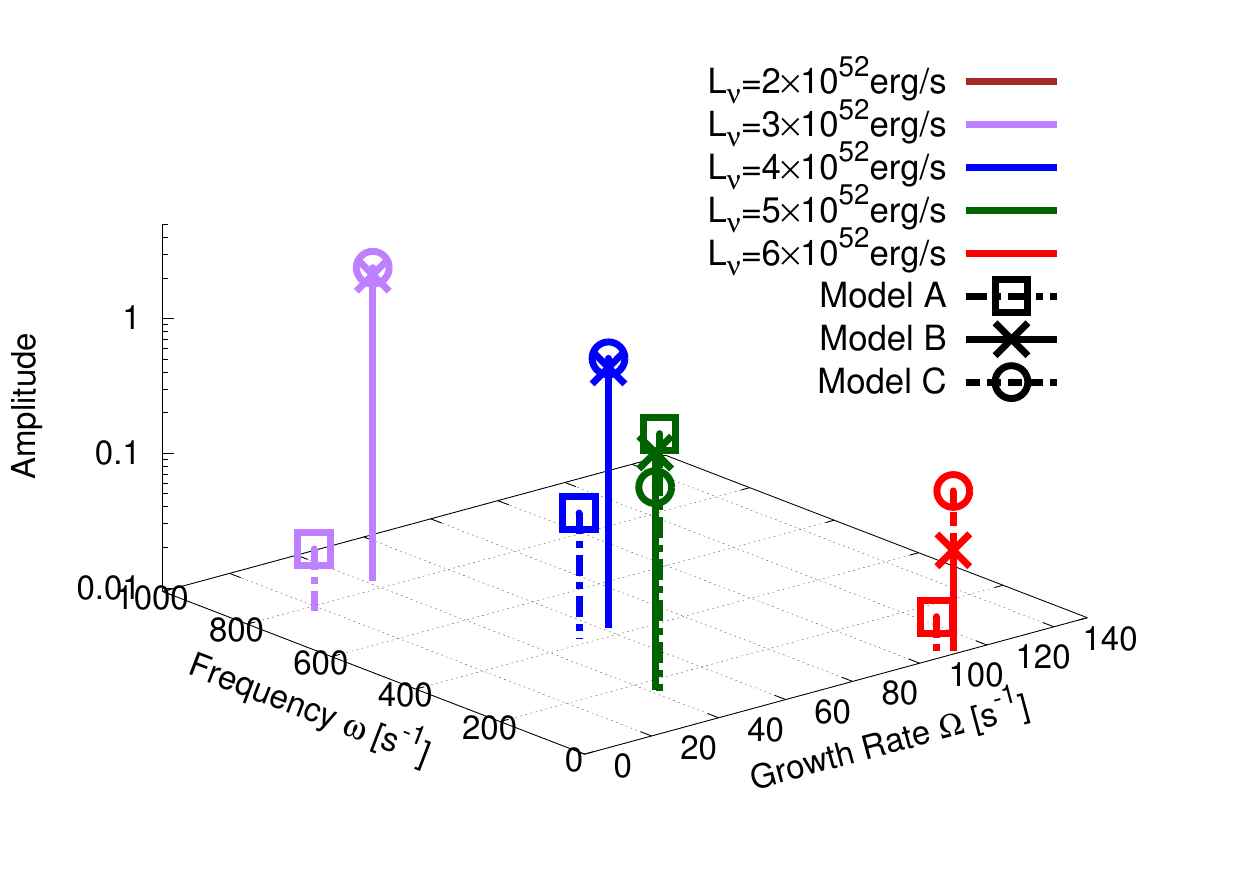}{0.5 \textwidth}{(a) $l = 1$}
 \fig{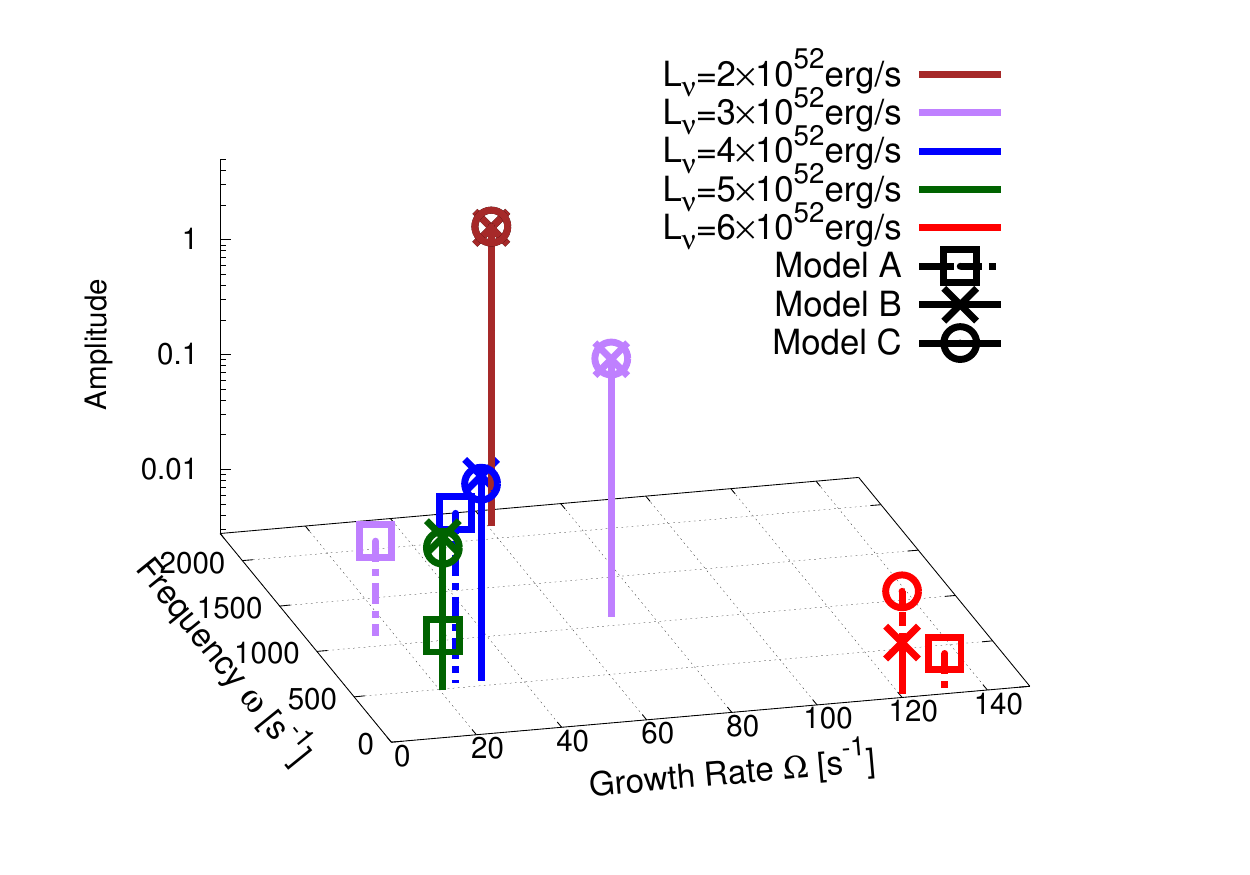}{0.5 \textwidth}{(b) $l = 2$}}
  \gridline{
 \fig{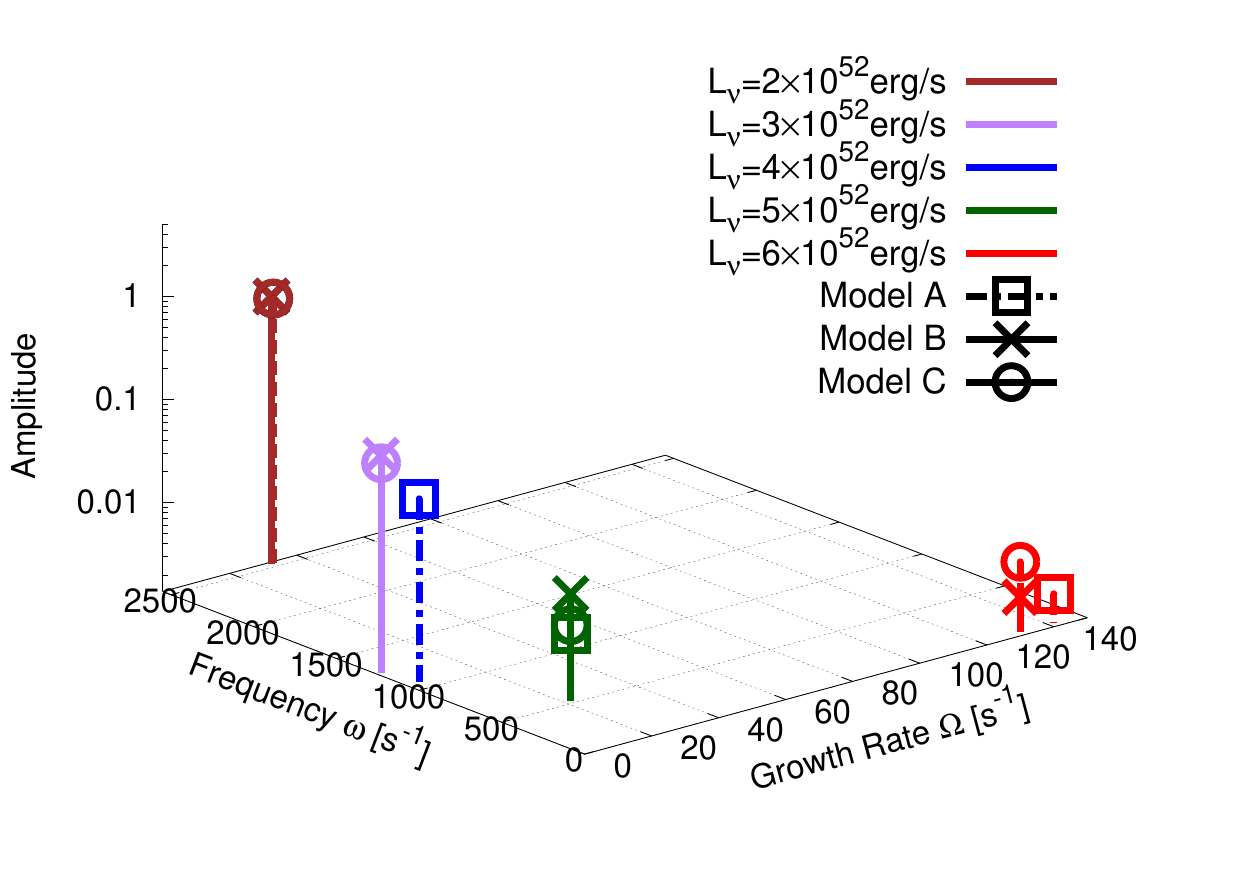}{0.5 \textwidth}{(c) $l = 3$}
 \fig{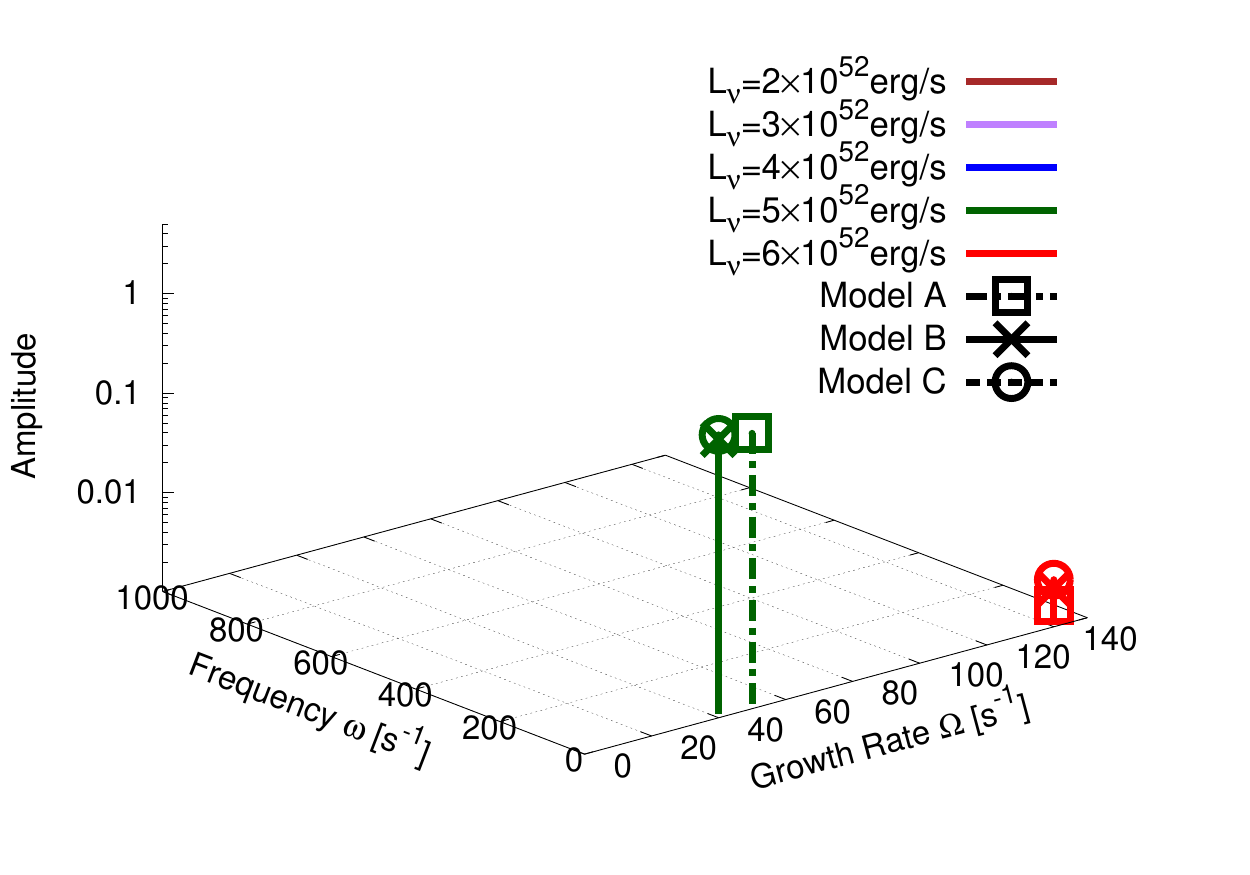}{0.5 \textwidth}{(d) $l = 4$}}
  \caption{Amplitudes of unstable eigenmodes for different boundary conditions, luminosities and spherical harmonics modes.  Dash-dotted lines and squares correspond to model A, solid lines and crosses mean model B and dashed lines and circles represent model C. Colors indicate the different neutrino luminosities.}
\label{fig:amplitude}
 \end{figure*}

As we mentioned earlier, the amplification of the mode amplitudes by the acoustic power injection becomes more remarkable as the neutrino luminosity gets lower. 
We discuss this trend from the view point of the resonance between SASI  and PNS g-modes. 

We omit the perturbation to the neutrino luminosity, that is, we assume $\bm{\tilde{L}}(s) = 0$, $\bm{w} = 0$ and $\delta T_{\alpha} = 0$ in equation (\ref{eq:GeneralSol}) in the rest of this subsection for simplicity. Then the Laplace-transformed shock perturbation is written as 
\begin{eqnarray}
 \dfrac{\delta r_{\mathrm{sh}}^*}{r_{\mathrm{sh}}} &=& \dfrac{\bm{a}^*(s)\cdot  \tilde{\Lambda}^*(s)R\bm{z}^*(s) +\bm{a}^*(s)\cdot \bm{\tilde{h }}^*[\bm{y}_0](s) -b^*(s)}{\bm{a}^* \! \cdot\tilde{\Lambda}^*(s) (s\bm{c}+\bm{d})} \notag \\
\\
&=&: \left( \dfrac{\delta r_{\mathrm{sh}}^*}{r_{\mathrm{sh}}} \right)_{\mathrm{ex}} + \left( \dfrac{\delta r_{\mathrm{sh}}^*}{r_{\mathrm{sh}}} \right)_{\mathrm{ini}} +\left( \dfrac{\delta r_{\mathrm{sh}}^*}{r_{\mathrm{sh}}} \right)_{\mathrm{IB}} ,
\end{eqnarray}
where we define 
\begin{eqnarray}
 \left( \dfrac{\delta r_{\mathrm{sh}}^*}{r_{\mathrm{sh}}} \right)_{\mathrm{ex}} &:= \dfrac{\bm{a}^*(s)\cdot  \tilde{\Lambda}^*(s)R\bm{z}^*(s)}{\bm{a}^* \! \cdot\tilde{\Lambda}^*(s) (s\bm{c}+\bm{d})}, \\
 \left( \dfrac{\delta r_{\mathrm{sh}}^*}{r_{\mathrm{sh}}} \right)_{\mathrm{ini}} &:= \dfrac{\bm{a}^*(s)\cdot \bm{\tilde{h }}^*[\bm{y}_0](s)}{\bm{a}^* \! \cdot\tilde{\Lambda}^*(s) (s\bm{c}+\bm{d})}, \\
 \left( \dfrac{\delta r_{\mathrm{sh}}^*}{r_{\mathrm{sh}}} \right)_{\mathrm{IB}} &:= \dfrac{-b^*(s)}{\bm{a}^* \! \cdot\tilde{\Lambda}^*(s) (s\bm{c}+\bm{d})}, \label{eq:IBterm} 
\end{eqnarray}
which represent the contributions from the perturbations imposed at the outer boundary, those set initially and those given at the inner boundary, respectively.

\begin{figure}[tbp!]
 \includegraphics[scale=0.23]{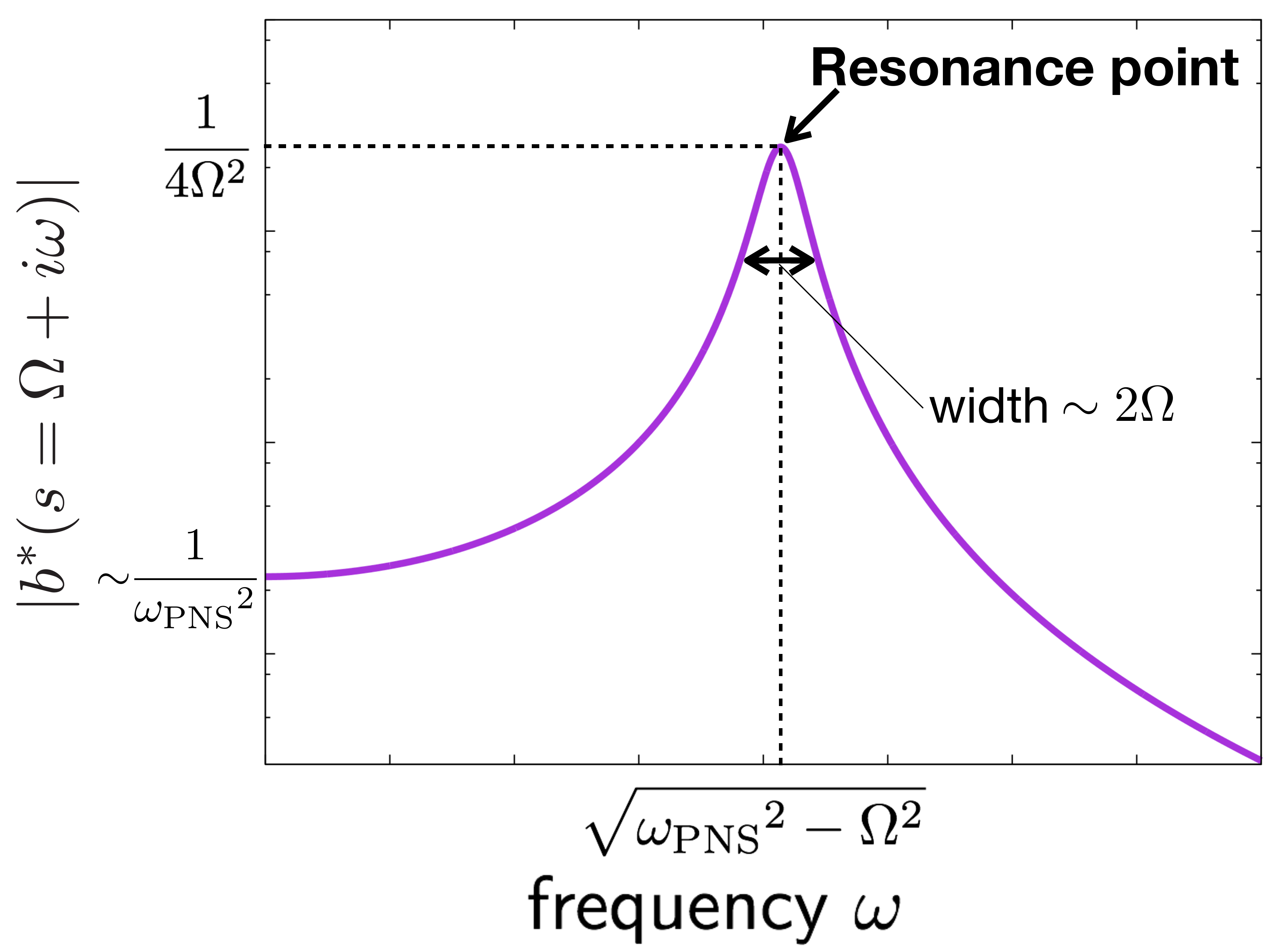}
  \caption{$|b^*(s = \Omega + i \omega)|$ as a function of $\omega$. The vertical axis is logarithmic scale. The maximum value of $|b^*|$ is obtained at $\omega = \sqrt{\omega_{\mathrm{PNS}}^2 - \Omega^2}$ where the acoustic wave produced by the g-mode oscillation of the PNS becomes resonant with the SASI oscillation.}
\label{fig:aib}
 \end{figure}

 We pay attention to the last one (equation (\ref{eq:IBterm})). The amplitude of an eigenmode is proportional to $b^*(s)$ at the pole corresponding to this mode. 
In the acoustic injection model considered here, $b(t)$ is assumed to be monochromatic and given as 
\begin{equation}
 b(t) = \sin \left( \omega_{\mathrm{PNS}} t \right), 
\end{equation} 
(see equation (\ref{eq.outgoing})) and its Laplace transform is then obtained as
\begin{equation}
 b^*(s) =  \dfrac{\omega_{\mathrm{PNS}}}{s^2 + {\omega_{\mathrm{PNS}}}^2} .
\end{equation}
The amplitude of eigenmode is hence proportional to 
\begin{equation}
 b^*(s = \Omega + i \omega) = \dfrac{\omega_{\mathrm{PNS}}}{\left(\Omega + i \omega \right)^2 + {\omega_{\mathrm{PNS}}}^2},
\end{equation}
and the maximum value of $|b^*|$ is given at $\omega = \sqrt{\omega_{\mathrm{PNS}}^2 - \Omega^2}$ 
(see Figure \ref{fig:aib}).
Considering the fact that  $\Omega^2 \ll \omega_{\mathrm{PNS}}^2$ in the typical situation, in which  $\omega_{\mathrm{PNS}} \gtsim 1000 \  \mathrm{s^{-1}}$ (see, for example, \citet{Burrows06}) and $\omega \sim 100 \ \mathrm{s^{-1}}$ (see Figure \ref{fig:pole12}), this is essentially $\omega \sim \omega_{\mathrm{PNS}}$ which means that the oscillation of the unstable mode should be resonant with one of the g-mode oscillations. 
Such a situation is approached as the neutrino luminosity is lowered, since the shock radius is decreased and, as a result, the SASI frequency is increased (see Figure \ref{fig:aib}).
This is essentially true irrespective of the value of $l$ as can be seen in the right panels of Figure \ref{fig:pole34}.


\begin{figure}[t!]
\epsscale{1.2}
\plotone{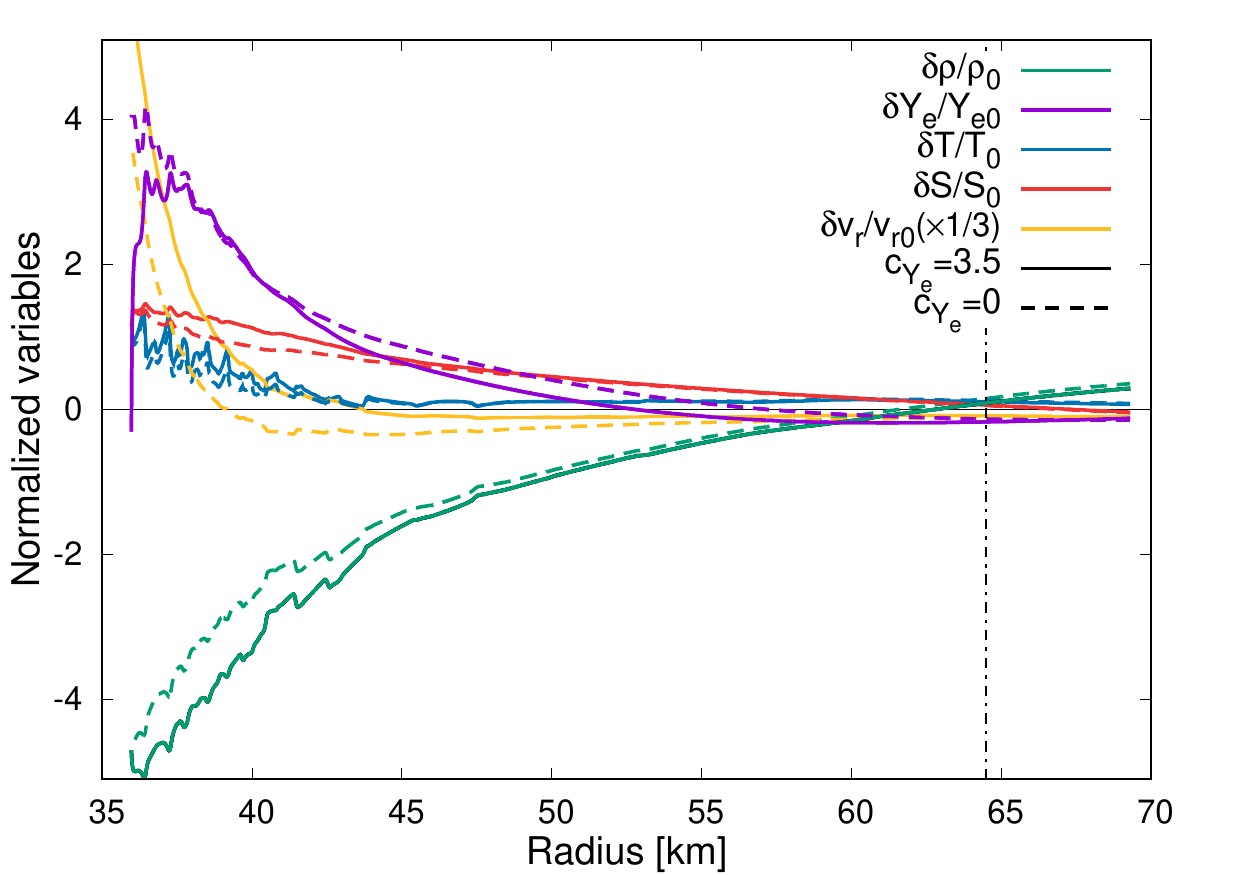}
  \caption{Radial distributions of various quantities in a steady perturbed solution. We set $L_{\nu} = 3.0 \times 10^{52} \ \mathrm{erg/s}$ and $l = 1$ mode here. Dashed and solid lines correspond to the models for $c_{Y_e} = 0 $ and $3.5$, respectively. Vertical dash-dotted line indicates the gain radius.}
\label{fig:thermovariables_L30_l1}
 \end{figure}

\begin{deluxetable*}{CCCC|CCC}
 \tablecaption{Steady perturbed solutions \label{tab:spp}}
 \tablehead{
   \colhead{} & \multicolumn{3}{c}{$c_{Y_e} = 0$} & \multicolumn{3}{|c}{$c_{Y_e} = 3.5$}\\
\cline{2-7}
\colhead{\shortstack[c]{Neutrino luminosity \\ ($\mathrm{erg \  s^{-1}}$)} } & $\delta r_{\rm{sh}}/r_{\rm{sh}0}$ & $\delta T_{\nu} / T_{\nu 0}$ & $\delta r_{\nu_{e}}/r_{\nu_{e} 0}$ & \multicolumn{1}{c}{$\delta r_{\rm{sh}}/r_{\rm{sh}0}$} & $\delta T_{\nu} / T_{\nu 0}$ & $\delta r_{\nu_{e}}/r_{\nu_e 0}$
}
 \startdata
 2 \E{52} & 2.04 \E{-2} & 1 & 0 & 7.66 \E{-3}  & 1 & -0.650\\
 3 \E{52} & 1.86 \E{-1} & 1 & 0 & 1.51 \E{-1}  & 1 & -0.533\\
 4 \E{52} & 4.81 \E{-1} & 1 & 0 & 4.43 \E{-1}  & 1 & -0.421 \\
 5 \E{52} & 1.63 & 1 & 0 & 1.60  & 1 & -0.219 \\
 \enddata
\end{deluxetable*}

\begin{figure}[t!]
\epsscale{1.18}
\plotone{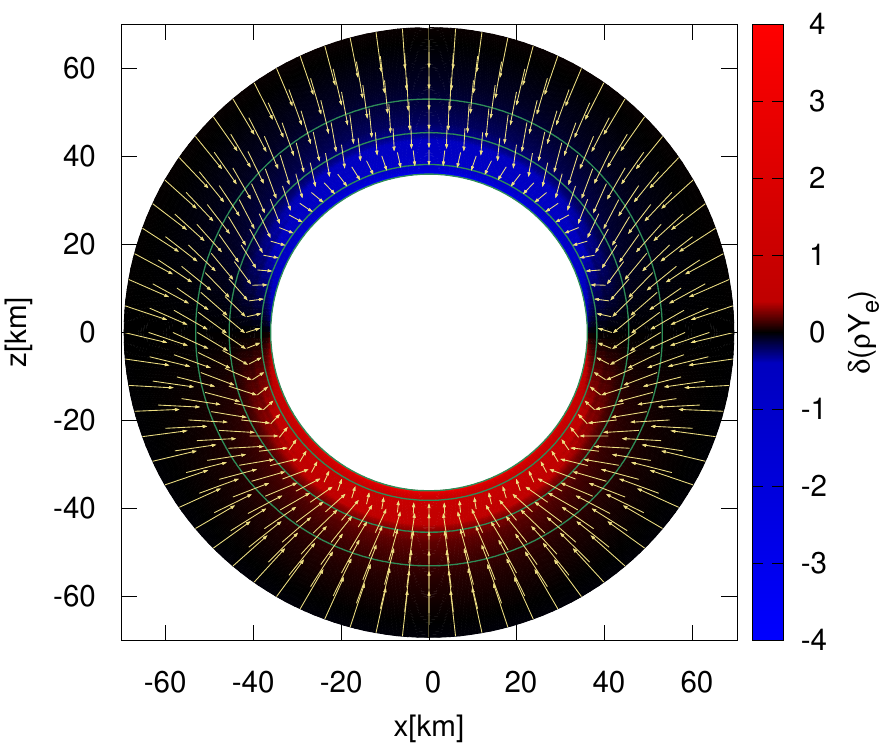}
  \caption{The distributions of velocity (arrows) and $\delta (\rho Y_e)$ (color map) in the meridional section for the model with $L_{\nu} = 3.0 \times 10^{52} \ \mathrm{erg/s}$. We set $l = 1$, $m = 0$ and $c_{Y_e} = 3.5$ here. The density contours are also shown for $ \rho_0 = 0.5, 1, 5$ and $10\E{10} \mathrm{g/cm^3}$. Note that the velocity perturbation is emphasized arbitrarily.}
\label{fig:flow_l1m0}
 \end{figure}

\newpage
\subsection{Steady perturbed state}
We now shift our attention to time-independent solutions of the perturbation equations.
This is motivated by the finding first reported by \citet{Tamborra14} that there occurs a shock deformation accompanied by an anisotropy in the number flux difference $F^n_{\nu_e} - F^n_{\bar{\nu}_e}$ of the electron-type neutrinos.
The structure is robust, being sustained for a long time. 
They called it the Lepton-number Emission Self-sustained Asymmetry (LESA).
\citet{Dolence15} later reported that they found a dipolar asymmetry not in the difference but in the sum of the two fluxes $F_{\nu_e} + F_{\bar{\nu}_e}$ (and also the sum of number fluxes) is strongly correlated with the deformation of the shock front. 
Regardless of the apparent discrepancy, these structures are supposed to be sustained by the combination of the shock deformation and the asymmetric neutrino emissions and may be produced even at the linear level. 
This is what we are going to investigate in this subsection.

For this purpose, we modify the basic equations as follows. We first extend the expressions for the fluctuations in the neutrino luminosities as
\begin{eqnarray}
 \dfrac{\delta L_{\nu_e}}{L_0} &= 4 \dfrac{\delta T_{\nu_e}}{T_{\nu_e 0}} + c_{Y_e} \dfrac{\delta Y_e}{Y_{e0}}  \label{eq.cYe_nue}, \\
 \dfrac{\delta L_{\bar{\nu}_e}}{L_0} &= 4 \dfrac{\delta T_{\bar{\nu}_e}}{T_{\bar{\nu}_e 0}} - c_{Y_e} \dfrac{\delta Y_e}{Y_{e0}}, \label{eq.cYe_nueb} 
\end{eqnarray}
where $c_{Y_e}$ is a constant parameter that accounts for the possible correlation between the fluctuation in the neutrino luminosity and that in $Y_e$ (\citet{Janka16}).
We investigate two cases with $c_{Y_e} = 0$ and $c_{Y_e} = 3.5$.
The latter value is taken from the numerical simulation of core-collapse supernova in 2D by \citet{Nagakura18} with Furusawa's EoS (\citet{Furusawa13}).
Since the fluctuations of neutrino luminosities can be also expressed as
\begin{eqnarray}
 \dfrac{\delta L_{\alpha}}{L_0} = 4 \dfrac{\delta T_{\alpha}}{T_{\alpha 0}} + 2 \dfrac{\delta r_{\alpha}}{r_{\alpha 0}} \label{eq.lba}
\end{eqnarray} 
with $\alpha = \nu_e , \bar{\nu}_e$, we obtain the perturbations to the neutrino spheres as follows:
\begin{eqnarray}
 \dfrac{\delta r_{\nu_e}}{r_{\nu_e 0}} &= + \dfrac{c_{Y_e}}{2} \dfrac{\delta Y_e}{Y_{e0}}, \\
 \dfrac{\delta r_{\bar{\nu}_e}}{r_{\bar{\nu}_e 0}} &= - \dfrac{c_{Y_e}}{2} \dfrac{\delta Y_e}{Y_{e0}}. 
\end{eqnarray}
We use these expressions just for numerical convenience. Note also that we assume
\begin{eqnarray}
  \dfrac{\delta T_{\alpha}}{T_{\alpha 0}} &= \dfrac{\delta T}{T_{0}},
\end{eqnarray}
that is, the fluctuations of the neutrino temperatures are equal to that of matter temperature at the neutrino sphere.

It is important to recognize that steady solutions of the linearized equations in the temporal regime correspond to the nontrivial solution at $s = 0$ of the Laplace-transformed equations. Since in this subsection we neglect both the acoustic injection from the inner boundary and the perturbation ahead of the shock front, such solution represent self-sustained non-spherical configurations, something corresponding to LESA at the linear perturbation level. Note that the existence of such solutions is itself highly non-trivial. 

It turns out that there are such solutions indeed even for $c_{Y_e} = 0$. As a typical case, we give the results for the model with $L_{\nu} = 3.0 \times 10^{52} \  \mathrm{erg \ s^{-1}}$ in the left half of Table \ref{tab:spp} and also as dashed lines in Figure \ref{fig:thermovariables_L30_l1}.
We consider only $l = 1$ modes here. Note also that $\delta T_{\alpha} / T_{\alpha 0}$ is normalized as unity.
Since we set  $c_{Y_e} = 0$, there is no difference in the fluctuation in the luminosity between $\nu_e$ and $\bar{\nu}_e$: 
\begin{equation}
 \delta L_{\nu_e} = \delta L_{\bar{\nu}_e}.
\end{equation}
As represented in Table \ref{tab:spp}, the deformation of the shock front $\delta r_{\rm{sh}}/r_{\rm{sh}}$ is correlated with the temperature fluctuation at the neutrino sphere, having the same sign.
This is consistent with the finding by \citet{Dolence15} although we cannot say anything about LESA in this case as there is no asymmetry in $\delta L_{\nu_e}$ and $\delta L_{\bar{\nu}_e} $ by definition.
It is noteworthy, however, that $\delta Y_e/ {Y_e}_0$ has the same signature as $\delta T/T_0$ on the PNS surface (see the purple and blue dashed lines in Figure~\ref{fig:thermovariables_L30_l1}).

Next we consider the case with $ \delta L_{\nu_e} - \delta L_{\bar{\nu}_e} \neq 0$, employing the models with $c_{Y_e} = 3.5$. 
The results are given in the right half of Table \ref{tab:spp} and also as solid lines in Figure \ref{fig:thermovariables_L30_l1}. We obtain non-trivial solutions also in this case.
The configurations are not much different from these for the $c_{Y_e} = 0$ case with $\delta r_{\rm{sh}}/r_{\rm{sh}}$ being positively correlated with the temperature fluctuation.
On the other hand, the perturbation to the electron fraction, $\delta Y_e/ {Y_e}_0$, has the opposite sign as $\delta T/T_0$ on the PNS surface (see the purple and blue solid lines in Figure \ref{fig:thermovariables_L30_l1}). 
As a consequence of the latter fact, the neutrino luminosity difference $\delta L_{\nu_e} - \delta L_{\bar{\nu}_e}$ becomes negatively correlated with the shock deformation.
We show the distributions of the velocity and the lepton number fluctuation in the meridional section in Figure \ref{fig:flow_l1m0}.
They are consistent with the schematic picture drawn by hand in \citet{Tamborra14} (see their Figure 15): 
the dipolar deformation of the shock front bends stream lines and induces the asymmetry in the lepton number near the PNS surface, which is anti-correlated with the shock deformation, i.e., more lepton-rich matter accretes on the hemisphere, in which the shock recedes, resulting in positive $\delta Y_e$ there (see the color map in Figure \ref{fig:flow_l1m0} for $\delta (\rho Y_e)$).
Note that in drawing the velocity field, we add to the unperturbed flow the perturbation scaled arbitrarily.

Looking into these configurations more closely, we find that $|v_r|$ decreases in the direction, in which the shock front expand (see the yellow lines in Figure \ref{fig:thermovariables_L30_l1}).
This results in longer heating in the gain region and the temperature gets higher on this side.
This then leads to larger luminosities on the same side. 
We also find that $Y_e$ becomes smaller at the original shock position (see the purple line in Figure \ref{fig:thermovariables_L30_l1}), since the shock expansion leads to an earlier turn-on of the electron capture. 
As the matter flows downward, $\delta Y_e$ increase owing to the lager neutrino luminosities just mentioned, which enhances the absorption of both $\nu_e$ and $\bar{\nu}_e$ with the former being dominant.
The behavior of $\delta Y_e$ near the PNS surface is qualitatively different between the models with and without the correlation of the perturbation of the neutrino luminosity and that of $Y_e$.
As a matter of fact, if we take it into account, adopting $c_{Y_e} = 3.5$, $\delta Y_e$ decreases rapidly and changes its sign. It is also observed that $\delta Y_e$ is initially smaller in this case than in the previous case with $c_{Y_e} = 0$.
This is because $F_{\bar{\nu}_e} > F_{\nu_e}$ is established in the direction of the deformation of the shock expansion, which in turn is produced by the asymmetry of $Y_e$ on the PNS surface mentioned above: $Y_e$ is depleted on the same side.
The decrease in $\delta Y_e$ and hence $Y_e$ itself there is induced by the \replaced{enhances}{enhanced} absorption of $\bar{\nu}_e$. These explanations are admittedly tautological because it is difficult to tell the cause from the result in the steady state and all we can say is that the configuration is self-consistent.
Incidentally, $\delta Y_e$ converges to the same value at $r \sim 40 \mathrm{km}$ irrespective of the value of $c_{Y_e}$. 
This is a well-known feedback effect:
the initial decrease of $\delta Y_e$ via the enhanced absorption is canceled by the suppression of electron capture later because of the smaller value of $Y_e$. 
It should be also mentioned that the very rapid change of $\delta Y_e$ near the inner boundary in the case of $c_{Y_e} = 3.5$ is an artifact of the geometric factor we employ in this study.
In fact, the perturbation to $\lambda$ (see equation (\ref{eq.lambda_nu})) is written as 
\begin{equation}
 \delta \lambda_{\nu} = \left( \dfrac{\partial \lambda_{\nu}}{\partial T_{\alpha}} \right) \delta T_{\alpha} +
  \left( \dfrac{\partial \lambda_{\nu}}{\partial Y_e} \right) \delta Y_e|_{\mathrm{PNS}}. 
\end{equation}
The second term on the right hand side is rewritten as
\begin{eqnarray}
 \left( \dfrac{\partial \lambda_{\nu}}{\partial Y_e} \right) \delta Y_e|_{\mathrm{PNS}} &&= \left( \dfrac{\partial \lambda_{\nu}}{\partial r_{\nu}} \right) \delta r_{\nu} \notag \\ 
&&\propto \dfrac{\partial g(r) }{\partial r_{\nu}} =\dfrac{1}{2} \dfrac{\left(r_{\nu}/r \right)^2}{\sqrt{1 - \left(r_{\nu}/r \right)^2}},  
\end{eqnarray}
in which the rightmost expression is a contribution from the derivative of the geometrical factor and is divergent at $r = r_{\nu}$.
In reality, $\delta \lambda$ and hence $\delta Y_e$ should change more gradually.

The fact that the self-sustained steady states are obtained irrespective of the value of $c_{Y_e}$ seems to suggest that the asymmetry of the lepton number flux is not essential for the build-up of such configurations but rather a structure associated with them.
Indeed our results appear to indicate that it is the temperature fluctuation and the resultant perturbation to the sum of the fluxes of $\nu_e$ and $\bar{\nu}_e$, $F_{\nu_e} + F_{\bar{\nu}_e}$ rather than their difference that is the most important ingredients for the production of the self-sustained configurations (\citet{Dolence15}). 
It is also true, however, that the asymmetries in the lepton fluxes and in the $Y_e$ distribution are also self-sustained and are consistent with what \citet{Tamborra14} observed in their simulations.
In this sense, our results are consistent with both of them in fact.

\section{Summary and conclusion} \label{sec:summary}
We have investigated the instability of the standing shock wave and the accretion flows downstream in the core of CCSNe by linear analysis. 
We have focused particularly on the influences of the fluctuations at the inner boundary.
As possible sources of such fluctuations, we have considered the injection of acoustic powers and the perturbation of the neutrino luminosity. 
We have linearized hydrodynamics equations and solved them by Laplace transform, regarding them as an initial-boundary-value problem.
As for the unperturbed state, we employ spherically symmetric, steady shocked accretion flows. 
In this formulation, the intrinsic eigenmodes such as SASI or convection are obtained as pole singularities of the Laplace-transformed equations and how strongly these modes are excited by the perturbations imposed at the boundaries is given as the residues at the poles.

We have first explored the influence of the injection of acoustic power from the PNS. 
Unlike the perturbation at the outer boundary, those at the inner boundary like this one change the growth rates and oscillation frequencies of eigenmodes directly.
We have hence calculated them together with the excitation amplitudes for various neutrino luminosities. 
We have found that the acoustic power injection enhances the growth rates of the fluid instabilities in general and the critical luminosity, i.e., the lowest luminosity at which SASI occurs for some modes, is reduced.
The enhancement is especially remarkable at low neutrino luminosities.
This is because the mismatch between the g-mode oscillations of the PNS and the oscillations in the accretion flow is reduced.
In fact, as the luminosity is lowered, the shock shrinks and, as a result, the latter frequency rises:
since the g-mode frequency is normally higher than the frequency of the oscillations in the accretion flow, this means that
the two frequencies get closer to each other.
We have observed that the perturbation from the inner boundary excites some modes more preferentially than others as the perturbation from the outer boundary does (see paper I) but the excitation amplitudes are also enhanced by the injection of acoustic powers.
On the other hand, the perturbations of the neutrino luminosities have been shown to have little effect on the growth rates and oscillation frequencies of eigenmodes as well as on their excitation amplitudes at least at the linear level.
\added{In this study, we have considered only the luminosity fluctuations of neutrinos originating from the neutrino sphere and ignored the emission from accreting matter in the cooling region. We do not think that their incorporation will change our conclusion on the effect of the fluctuations in the neutrino luminosity on the shock instability, since they will behave in a similar way to the core neutrinos.}

We have then investigated steady perturbed solutions, being motivated by the LESA phenomenon (\citet{Tamborra14}).
We have turned off the perturbations on the outer and inner boundaries except for those in the neutrino luminosities, which we assumed are induced by the fluctuation in temperature and/or $Y_e$, and studied if self-sustained configurations are obtained or not.
We have found solutions indeed in both cases.
It should be stressed that it is not a trivial thing.
We have shown that the obtained structures are consistent with that proposed by \citet{Tamborra14}.
We also found that such self-sustained steady states are obtained without asymmetry between $\nu_e$ and $\bar{\nu}_e$ and it seems that it is the temperature fluctuation and the resultant perturbation to the sum of the fluxes of $\nu_e$ and $\bar{\nu}_e$, $F_{\nu_e} + F_{\bar{\nu}_e}$, rather than their difference that is the key to the production of these steady states.
This seems consistent with the claim by \citet{Dolence15}.
On the other hand, the asymmetries in the lepton emissions and in the $Y_e$ distribution as \citet{Tamborra14} found them are also self-sustained although they may be a by-product.
In this sense our results appear to be consistent also with their results.

The models we employed in this paper are admittedly a very crude approximation to reality.
\deleted{In fact, the use of very simple EoS and the neglect of dissociation of nuclei by shock heating give systematically higher temperature in the post-shock flow.}
\added{Neutrino transfer is neglected and is replaced with the light bulb approximation.}
It is known that the geometrical factor adopted in this paper tends to give larger critical luminosities.
The numbers obtained for these quantities in this paper should not be taken at face value.
Recently our group has succeeded to obtain the axisymmetric steady states with rotation and/or magnetic field (\citet{Fujisawa18}).
It will be hence interesting to extend the linear stability analysis to these configurations.

\acknowledgments

We thank W. Iwakami, A. Harada and K. Nakmura for providing us the data of their numerical simulations of CCSNe and also for helpful discussions. This work is partially supported by the Grant-in-Aid for the Scientific Research from the Ministry of Education, Culture, Sports, Science and Technology (MEXT), Japan (16H03986).

\clearpage
\appendix
\twocolumngrid
\section{The Explicit form of $\bm{u}$ in equation (47)} \label{sec:v}
The vector $\bm{y}$ for the perturbed quantities is defined in equation (\ref{eq:perturbedstate}) and its basic equations are schematically given first as 
\begin{equation}
 M \dfrac{\partial \bm{y}}{\partial t} + A' \dfrac{\partial \bm{y}}{\partial r} + B' \bm{y} = \bm{u}',
\end{equation}
where matrices $M(r)$, $A'(r)$ and $B'(r)$ are the same as those given in equations (49), (50) and (51) of appendix A in paper I, respectively, and
\begin{equation}
 \bm{u}' = \left(0, 0, 0, \dfrac{1}{v_r} \delta q_{\nu}, \dfrac{m_b}{\rho Y_e} \delta \lambda_{\nu}, 0\right)^{T}.
\end{equation}
We then obtain equation (\ref{eq.basicLESA}) by defining $A := - A'^{-1} M$, $B := -A'^{-1} B$ and
\begin{equation}
 \bm{u} \dfrac{\delta T_{\alpha}}{T_{\alpha}} := A'^{-1} \bm{u}'.
\end{equation}

\section{The Formal solution} \label{sec:formal sol.}
We present here the formal solution of equation (\ref{eq.leqLESA}), which includes the acoustic power injection as well as the fluctuation of neutrino luminosity.
The equation, 
\begin{eqnarray}
\label{An0}
\frac{\diff \bm{y}^*}{\diff r} &=& (sA +B)\bm{y}^* -A\bm{y}_0(r) + \bm{u} \left(\dfrac{\delta T_{\alpha}}{T_{\alpha}}\right)^*,
\end{eqnarray}
is integrated as 
\begin{eqnarray}
&&\bm{y}^*(r,s) \notag \\
&& = \Lambda^*(r,s) \bm{y}^* (r_\mathrm{sh}, s) -\Lambda^*(r,s)\int _{r_\mathrm{sh}}^r \diff r' \Lambda^{*-1}(r',s)A(r')\bm{y}_0(r') \notag \\
&& \ \ \ + \Lambda^*(r,s)\int _{r_\mathrm{sh}}^r \diff r' \Lambda^{*-1}(r',s) \bm{u} \left(\dfrac{\delta T_{\alpha}}{T_{\alpha}}\right)^*,
\\
\label{EqLast}
&& = \Lambda^*(r,s) \bm{y}^* (r_\mathrm{sh}, s) - \bm{h}^*[\bm{y}_0](r,s) + \bm{L}^*(r,s) \left(\dfrac{\delta T_{\alpha}}{T_{\alpha}}\right)^*,
\end{eqnarray}
where the matrix $\Lambda^*$ and vector $\bm{h}^*[\bm{y}_0]$ and $\bm{L}^*$ are defined as
\begin{eqnarray}
 &\Lambda^*(r,s) :=\mathcal{P} \left[ \exp \left( \displaystyle\int _{r_\mathrm{sh}} ^r \diff r' (sA+B) \right) \right], & \\
&\bm{h}^*[\bm{y}_0](r,s) := \Lambda^*(r,s) \displaystyle\int _{r_\mathrm{sh}}^r \diff r' \Lambda^{*-1}(r',s)A(r')\bm{y}_0(r'),& \notag \\
\\
&\bm{L}^*(r,s) := \Lambda^*(r,s) \displaystyle\int _{r_\mathrm{sh}}^r \diff r' \Lambda^{*-1}(r',s) \bm{u}(r').&
\end{eqnarray}
In the above expressions, $\mathcal{P}$ stands for the path-ordering operator (\citet{PeskinQFT}).
Inserting the linearized Rankine-Hugoniot relation (\ref{OB}), we evaluate equation (\ref{EqLast}) at the neutrino sphere as
\begin{eqnarray}
 \bm{y}^*(r_{\nu _e},s) &=& \Lambda^*(r_{\nu _e},s) \left[ (s\bm{c}+\bm{d}) \frac{\delta r_{\mathrm{sh}}^*(s)}{r_\mathrm{sh}} + R\bm{z}^*(s) \right] \notag \\
                         &&- \bm{h}^*[\bm{y}_0](r_{\nu _e},s) + \bm{L}^*(r_{\nu}, s) \left(\dfrac{\delta T_{\alpha}}{T_\alpha} \right)^{*}(s) ,\\
\label{An1} 
&=: &\tilde{\Lambda}^*(s) \left[(s\bm{c}+\bm{d}) \frac{\delta r_\mathrm{sh}^*(s)}{r_\mathrm{sh}} +R\bm{z}^*(s) \right] \notag \\
   &&- \bm{ \tilde {h}}^*[\bm{y}_0](s) + \bm{\tilde{L}}^*(s) \left( \dfrac{\delta T_{\alpha}}{T_{\alpha}} \right)^*(s),
\end{eqnarray}
in which $\tilde{\Lambda}^*(s)$, $\bm{ \tilde {h}}^*[\bm{y}_0](s)$ and $\bm{\tilde{L}}^*(s)$ are the values of $\Lambda^*$, $\bm{h}^*[\bm{y}_0]$ and  $\bm{L}$ evaluated at the neutrino sphere, which are introduced just for notational simplicity.

The boundary conditions imposed at $r_{\nu_e}$ are generally written as
\begin{eqnarray}
&\bm{a}^*(s)\cdot \bm{y}^* (r_{\nu _e},s) + b^*(s)= 0,& \\
& \left(\dfrac{\delta T_{\alpha}}{T_{\alpha}}\right)^*(s) = \bm{w}(s) \cdot \bm{y}^* (r_{\nu _e},s), \label{An2}&
\end{eqnarray}
the latter of which corresponds to equation (\ref{eq.IBLESA}).
Substituting the formal solution into these two equations, we obtain the followings:
\begin{eqnarray}
&& \left(\begin{array}{cc}
  \bm{a}^* \! \cdot\tilde{\Lambda}^*(s) (s\bm{c}+\bm{d}) \ \  &\  \bm{a}^* \! \cdot\bm{\tilde{L}}^*(s) \\
  \bm{w}^* \! \cdot\tilde{\Lambda}^*(s) (s\bm{c}+\bm{d}) \ \ & \ \bm{w}^* \! \cdot\bm{\tilde{L}}^*(s) - 1 
 \end{array} \right) 
\left(\begin{array}{c}
 \delta r_{\mathrm{sh}}^*/r_{\mathrm{sh}}\\
 \delta T_{\alpha}^* / T_{\alpha}
\end{array} \right)
  \notag \\
&&+
 \left(\begin{array}{c}
 \bm{a}^*(s)\cdot  \tilde{\Lambda}^*(s)R\bm{z}^*(s) - \bm{a}^*(s)\cdot \bm{\tilde{h}}^*[\bm{y}_0](s) + b^*(s) \\
 \bm{w}^*(s)\cdot  \tilde{\Lambda}^*(s)R\bm{z}^*(s) - \bm{w}^*(s)\cdot \bm{\tilde{h }}^*[\bm{y}_0](s)
\end{array} \right) \notag \\
&&= 0. \label{eq:GeneralSol}
\end{eqnarray} 
We find from these equations the positions of poles as the zeros of 
\begin{equation}
 \left|\begin{array}{cc}
  \bm{a}^* \! \cdot\tilde{\Lambda}^*(s) (s\bm{c}+\bm{d}) \ \  &\  \bm{a}^* \! \cdot\bm{\tilde{L}}^*(s) \\
  \bm{w}^* \! \cdot\tilde{\Lambda}^*(s) (s\bm{c}+\bm{d}) \ \ & \ \bm{w}^* \! \cdot\bm{\tilde{L}}^*(s) - 1 
 \end{array}\right| = 0. \label{eq:det}
\end{equation}
Note that they are not affected by the upstream perturbation $\bm{z}^* (s)$ but are indeed influenced by the inner boundary conditions, i.e., $\bm{a}^*(s)$ and $\bm{w}^*(s)$.

\bibliographystyle{aasjournal}

\end{document}